\documentclass[showpacs,amsmath,amssymb,twocolumn,prl,superscriptaddress,notitlepage]{revtex4-1}

\usepackage{algorithm}
\usepackage{algorithmic}
\usepackage{amsmath,amssymb,amsthm,mathrsfs,amsfonts,dsfont}
\usepackage{amsmath,bm}
\usepackage{array,booktabs}
\usepackage{bm}
\usepackage{braket}
\usepackage{cases}
\usepackage{color}
\usepackage{comment}
\usepackage{enumerate}
\usepackage{float}
\usepackage{framed}
\usepackage[top=2cm, bottom=2cm, left=2cm, right=2cm]{geometry}
\usepackage{graphicx}
\usepackage{here}
\usepackage{mathtools}
\usepackage[version=3]{mhchem}
\usepackage{pifont}
\usepackage{qcircuit}
\usepackage{subfigure, epsfig}
\usepackage{wrapfig}

\graphicspath{{./image/}}

\newcommand{\n}{\nonumber \\ }

\begin{document}


\title{Universal Scaling Bounds on a Quantum Heat Current}


\author{Shunsuke Kamimura}
\email{s2130043@s.tsukuba.ac.jp}
\affiliation{Faculty of Pure and Applied Sciences, University of Tsukuba, Tsukuba 305-8571, Japan}
\affiliation{Research Center for Emerging Computing Technologies,  National  Institute  of  Advanced  Industrial  Science  and  Technology  (AIST),1-1-1  Umezono,  Tsukuba,  Ibaraki  305-8568,  Japan.}

\author{Kyo Yoshida}
\affiliation{Faculty of Pure and Applied Sciences, University of Tsukuba, Tsukuba 305-8571, Japan}

\author{Yasuhiro Tokura}
\email{tokura.yasuhiro.ft@u.tsukuba.ac.jp}
\affiliation{Faculty of Pure and Applied Sciences, University of Tsukuba, Tsukuba 305-8571, Japan}

\author{Yuichiro Matsuzaki}
\email{matsuzaki.yuichiro@aist.go.jp}
\affiliation{Research Center for Emerging Computing Technologies,  National  Institute  of  Advanced  Industrial  Science  and  Technology  (AIST),1-1-1  Umezono,  Tsukuba,  Ibaraki  305-8568,  Japan.}


\begin{abstract}

In this Letter, we derive new bounds on a heat current flowing into a quantum $L$-particle system coupled with a Markovian environment.
By assuming that a system Hamiltonian and a system-environment interaction Hamiltonian are extensive in $L$,
we prove that the absolute value of the heat current scales at most as $\Theta (L^3)$ in a limit of large $L$.
Furthermore, we present an example of non-interacting particles globally coupled with a thermal bath, for which this bound is saturated in terms of scaling.
However, the construction of such a system requires many-body interactions induced by the environment,
which may be difficult to realize with the existing technology.
To consider more feasible cases,
we consider a class of the system where any non-diagonal elements of the noise operator
(derived from the system-environment interaction Hamiltonian)
become zero in the system energy basis,
if the energy difference exceeds a certain value $\Delta E$.
Then, for $\Delta E = \Theta (L^0)$,
we derive another scaling bound $\Theta (L^2)$ on the absolute value of the heat current,
and the so-called superradiance belongs to a class for which this bound is saturated.
Our results are useful for evaluating the best achievable performance of quantum-enhanced thermodynamic devices, including far-reaching applications such as quantum heat engines,
quantum refrigerators and quantum batteries.

\end{abstract}

\maketitle

Quantum mechanics successfully describes the counter-intuitive behaviors of microscopic objects, which cannot be explained by classical theory.
Recently, considerable effort has been devoted toward engineering large quantum systems
while sustaining quantum effects such as entanglement and coherence.
Quantum technology is a new field to seek novel industrial applications utilizing such quantum properties.

To quantify the performance of quantum devices,
we typically consider its scaling in a limit of a large number of qubits,
and we compare quantum and classical methods with equal resources
(e.g., using the same amount of time).
Shor demonstrated that a fault-tolerant quantum computer can solve factorization problems exponentially faster than the best-known classical algorithm~\cite{newshor1994algorithms}.
Quantum metrology exploits the non-classical properties of probe qubits to measure external fields with higher sensitivity.
The estimation uncertainty is known to decrease by $L^{-1/2}$ with $L$ classical probes;
in principle, the uncertainty can decrease by $L^{-1}$ with $L$ entangled qubits~\cite{huelga1997improvement,giovannetti2011advances,degen2017quantum}.

Meanwhile, since the Industrial Revolution,
thermodynamics has been conventionally adopted to describe the macroscopic behaviors of classical systems.
Its extension to fluctuating non-equilibrium classical small systems,
referred to as stochastic thermodynamics, has been studied extensively in recent decades~\cite{seifert2012stochastic}.
Specifically, stochastic thermodynamics can provide a richer set of limitations on heat engine performances,
as exemplified by fluctuation theorems~\cite{jarzynski1997nonequilibrium, crooks1999entropy, hatano2001steady, seifert2005entropy}
and trade-off relations~\cite{shiraishi2016universal, pietzonka2018universal}.

Quantum thermodynamics refers to the extension of thermodynamics to quantum systems~\cite{vinjanampathy2016quantum,binder2018thermodynamics,deffner2019quantum,auffeves2022quantum}.
Quantum versions of heat engines~\cite{alicki1979quantum, quan2007quantum,talkner2007fluctuation}
and batteries~\cite{alicki2013entanglement} have been proposed,
where the device is typically modeled as an open quantum system.
Several experimental demonstrations of such quantum thermodynamic devices have been reported~\cite{bergenfeldt2014hybrid, zhang2014quantum, pekola2015towards, altintas2015rabi, peterson2019experimental}.
As with other quantum technologies,
scaling advantages in the performances of heat engines~\cite{hardal2015superradiant,niedenzu2018cooperative, kloc2019collective,watanabe2020quantum,tajima2021superconducting, kloc2021superradiant,kamimura2022quantum,yadin2022thermodynamics,da2022collective,jaseem2023quadratic}
and quantum batteries~\cite{campaioli2017enhancing,ferraro2018high,tacchino2020charging,ito2020collectively,quach2022superabsorption,mayo2022collective,ueki2022quantum}
have been demonstrated for specific models.
Pioneered in the seminal work for the discovery of superradiance~\cite{dicke1954coherence},
many previous studies regarding a
scaling enhancement (with the system size)
of a heat current have been reported,
which plays a critical role in the advantage of quantum thermodynamic devices.
Despite recent discoveries such as a quadratic heat current in a high-temperature environment~\cite{hardal2015superradiant,niedenzu2018cooperative} or
interacting particles~\cite{kloc2019collective,kamimura2022quantum}
and more theoretical studies behind these systems~\cite{tajima2021superconducting,yadin2022thermodynamics},
universal scaling bounds on a heat current for arbitrary open quantum systems remain unknown.

\begin{figure}
	\begin{center}
		\includegraphics[clip,width=8cm,bb=0 0 350 175]{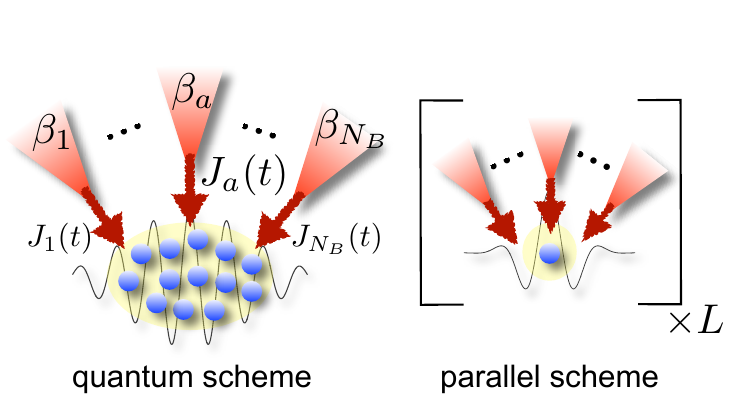}
		\caption{Schematic for a quantum scheme for heat current generation realized in a system composed of $L$ particles surrounded by $N_B$ baths.
		For a parallel scheme, the $L$ particles are used in parallel.}
		\label{fig:schematic}
	\end{center}
\end{figure}

In this Letter, we derive new bounds on a heat current $J (t)$ flowing into an open quantum system.
We consider an $L$-particle system whose Hamiltonian is extensive with respect to $L$.
A straightforward approach for generating the heat current is
the parallel use of $L$ particles,
where each particle is individually coupled with the environment.
We refer to this approach as a parallel scheme.
For this scheme, the heat current scales as $| J (t) | = \Theta (L) .$
Throughout this Letter, a function $f (L)$ is written as
$f (L) = \Theta \left( g (L) \right)$ 
if there exist constants $k_1, k_2$ and $L_0$, such that
$ k_1 g (L) \le f (L) \le k_2 g (L)$ is satisfied for any 
$L \ge L_0$.
Meanwhile, in a quantum scheme,
the particles are collectively coupled with the environment
(see Fig.~\ref{fig:schematic}).
For fair comparison of a quantum scheme and a parallel scheme,
we assume that a system-environment interaction Hamiltonian is extensive as well.
According to the convention of quantum thermodynamics,
we specifically focus on an open quantum system that is weakly coupled with the environment.
Under these conditions, we derive a scaling bound $|J (t)| \le \Theta (L^3)$,
and we present an example for which it is saturated in terms of scaling.
However, the model requires an $L$-body interaction,
which may be difficult to realize with the existing technology.
Therefore, to derive another bound for more feasible models,
we restrict the class of the system such that the interaction with the environment
only induces transitions between system energy eigenstates
whose energy difference is smaller than a certain value $\Delta E$.
By mathematically imposing this condition, we obtain another scaling bound
$|J (t)| \le \Theta (L^2)$ for $\Delta E = \Theta (L^0)$.
Furthermore, we find that in the case of superradiance,
which is a well-established collective energy emission process observed in
light-matter systems~\cite{dicke1954coherence},
this bound is saturated in terms of scaling.
Our derived bounds universally limit how a heat current can scale with the system-size,
regardless of the choice of the system.
Our results are useful for evaluating the best achievable performance of quantum-enhanced thermodynamic devices.
As an example, we can also construct a new quantum heat engine
with a quantum enhancement.


{\it{Model.}}\textemdash
We consider a total system composed of a quantum system subject to an environment modeled as $N_B$ heat baths.
The total Hamiltonian is expressed as follows:
\begin{align}
    \hat{H}_{\text{tot}} &= \hat{H}_S + \hat{H}_B + \hat{H}_{\text{int}},\\
    \hat{H}_B &= \sum_{a=1}^{N_B} \hat{H}^{(a)}_{B},\\
    \hat{H}_{\text{int}} &= \sum_{a=1}^{N_B} \hat{H}^{(a)}_{\text{int}},
\end{align}
where $\hat{H}_S$ ($\hat{H}_B$) is a system (an environmental) Hamiltonian,
$\hat{H}^{(a)}_{B}$ is a free Hamiltonian of the $a$-th bath,
and $\hat{H}_{\text{int}}$ is a system-environment
interaction Hamiltonian that is defined as
a summation of interaction Hamiltonians
$\hat{H}^{(a)}_{\text{int}}$
between the system and the $a$-th bath.
Further, we assume that these Hamiltonians are time-independent.
Let
$\hat{\rho}^{(a)}_{B} = e^{ -\beta_a \hat{H}^{(a)}_{B} }/\text{Tr}_a [ e^{ -\beta_a \hat{H}^{(a)}_{B} } ]$
denote a thermal equilibrium state of the $a$-th bath
with an inverse temperature $\beta_a$,
and let $\hat{\rho}_B = \bigotimes_{a=1}^{N_B} \hat{\rho}^{(a)}_{B} $ denote a product state of 
these thermal equilibrium states. 
Here, $\text{Tr}_a$ is a partial trace over the degrees of freedom of
the $a$-th bath.
For the interaction Hamiltonian $\hat{H}^{(a)}_{\text{int}}$,
we introduce Hermitian noise (bath) operators
$\{ \hat{A}^{(a)}_{k} \}_{k = 1, 2, \ldots , c_a }$
($\{ \hat{B}^{(a)}_{k} \}_{k = 1, 2, \ldots , c_a }$)
acting on the degrees of freedom of the system ($a$-th bath),
where the label $k$ denotes a {\it channel}
through which the system-bath interaction is induced,
and $c_a$ denotes the number of channels.
Then, the interaction Hamiltonian can be expressed as follows:
\begin{align}
    \hat{H}^{(a)}_{\text{int}} = \sum_{k=1}^{ c_a } \hat{A}^{(a)}_{k} \otimes \hat{B}^{(a)}_{k}.
    \label{eq:Hint_explicit}
\end{align}

After applying the Born and Markov approximations,
we obtain the Redfield equation as follows:
($\hbar = 1$)~\cite{breuer2002theory}:
\begin{align}
   \frac{d \hat{\rho}_S (t)}{dt} &= -i [ \hat{H}_S , \hat{\rho}_S (t) ] + \sum_{a=1}^{N_B} \mathcal{D}_a [\hat{\rho}_S (t)],
    \label{eq:Redfield}\\
        \mathcal{D}_a [\hat{\rho}] &= \int_{0}^{\infty} d s \sum_{k,l=1}^{c_a} C^{(a)}_{kl} (s)  \bigl[ \tilde{A}^{(a)}_{l} (-s) \hat{\rho} \hat{A}^{(a)}_{k} \n
    & - \hat{A}^{(a)}_{k} \tilde{A}^{(a)}_{l} (-s) \hat{\rho} \bigr] 
    + \text{h.c.} ,
\end{align}
where $\hat{\rho}_S (t)$ denotes a reduced density operator of the system (in the Schr\"{o}dinger picture),
and $\mathcal{D}_a$ denotes a dissipator due to the $a$-th bath.
Here, we define a correlation function as
$C^{(a)}_{kl} (s) = \text{Tr}_a [ \tilde{B}^{(a)}_{k} (s) \hat{B}^{(a)}_{l} \hat{\rho}^{(a)}_B ]$,
where an operator is transformed as $\hat{O} \mapsto \tilde{O} (t) = e^{ i \hat{H}_0 t } \hat{O} e^{ - i \hat{H}_0 t }$
with respect to the free Hamiltonian $\hat{H}_0 = \hat{H}_S + \hat{H}_B$.
According to the convention of quantum thermodynamics,
we define an instantaneous net heat current $J (t)$
from the environment to the system as follows:
\begin{align}
    J (t) = \text{Tr} \left( \hat{H}_S  \frac{d \hat{\rho}_S (t)}{dt} \right).
\end{align}
For a multi-bath environment, we obtain
an {\it additivity} $J (t) = \sum_{a=1}^{N_B} J_a (t)$,
where a heat current $J_a (t)$ from the $a$-th bath to the system is defined as follows:
\begin{align}
  J_a (t) = \text{Tr} \left( \hat{H}_S \mathcal{D}_a [ \hat{\rho}_S (t) ] \right) .
\end{align}

{\it{Bound 1.}}\textemdash
We can derive a general upper bound on the absolute value of the heat current as follows~\footnote{See the subsection II~A of Supplementary Material,
which includes Refs.~\cite{lenard1978thermodynamical,pedersen2019native,butler2011polarization,purcell1946resonance,wood2014cavity,bienfait2016reaching,agarwal1971brownian,houck2007generating},
for a detailed derivation of Bound 1.}:
\begin{align}
    | J_a (t) | &\le 4 \| \hat{H}_S \| \sum_{k,l=1}^{c_a}
   \Xi_{kl}^{(a)}
    \| \hat{A}_k^{(a)} \| \| \hat{A}_l^{(a)} \|,
    \label{eq:bound1}
\end{align}
where we introduce a coefficient
$\Xi_{kl}^{(a)} = \int_0^{\infty} ds | C_{kl}^{(a)} (s) |$
and an operator norm for a given operator $\hat{O}$
(induced by a vector norm $\| \ket{\psi} \| = \sqrt{ \braket{ \psi | \psi } }$),
which is expressed as
$\| \hat{O} \| = \sup_{\ket{\psi} } \frac{ \| \hat{O} \ket{\psi} \| }{ \| \ket{\psi} \| }$.
This newly derived bound is applicable to an arbitrary open quantum system,
as far as the Born and Markov approximations are validated.

{\it{Application of Bound 1 to $L$-particle systems.}}\textemdash
We apply the general bound~(\ref{eq:bound1}) to a system composed of (generally interacting) $L$ identical particles
(such as an $L$-qubit system) subject to $N_B$ heat baths,
and we analyze how a heat current scales with $L$
(see Fig.~\ref{fig:schematic}).
Suppose that the form of the system-environment interaction
is given as Eq.~(\ref{eq:Hint_explicit}).
In particular, we focus on the case
in which the system Hamiltonian is extensive in $L$, i.e. $\| \hat{H}_S \| = \Theta (L)$.
Moreover, we assume that the interaction Hamiltonian $\hat{H}_{\text{int}}$
is extensive as well,
i.e. $\| \hat{H}_{\text{int}} \| = \Theta (L)$,
which is a natural assumption for a fair comparison of the heat current with that for a parallel scheme.
Hence, we assume $\| \hat{A}^{(a)}_k \| = \Theta (L/c_a)$,
which means that the distribution of the interaction energy over each channel $k$ is homogeneous.
Furthermore, we assume that $\| \hat{B}^{(a)}_k \| = \Theta (L^0)$,
$\Xi_{kl}^{(a)} = \Theta (L^0)$
and $N_B=\Theta (L^0)$,
because these quantities are
solely determined by the properties of the environment.
Consequently, we show that
$|J_a (t)|$ and
$\sum_{a=1}^{N_B} |J_a (t)|$
scale at most as $\Theta (L^3)$.
Remarkably, this scaling exceeds
that of the superradiance~\cite{dicke1954coherence},
which is known as a collective emission of photons from $L$ qubits.

In the derivation explained above, we assume that the norm of the system-bath interaction scales linearly with the system size.
This assumption is valid for example when a spin ensemble collectively interacts with a cavity 
where the wavelength of the cavity photon is much larger than the distance between the spins~\cite{imamouglu2009cavity,kakuyanagi2016observation}.
However, when we consider the thermodynamic limit, it might be difficult
for the system-coupling strength
to scale linearly with the system size~\cite{hepp1973equilibrium,kirton2019introduction}.
Thus, we consider the case in which the system is not large enough to consider the thermodynamic limit but is 
large enough to observe the collective coupling between the system and environment.
For example, the collective effect
between a cavity and thousands of superconducting qubits was experimentally observed~\cite{kakuyanagi2016observation}.
Therefore, our method could be realized up to thousands of qubits.

Here, we present an example of an $L$-qubit system for which
the scaling bound $| J_a (t) | \le \Theta (L^3)$ obtained from Eq.~(\ref{eq:bound1}) is saturated.
We introduce a system Hamiltonian as follows:
\begin{align}
    \hat{H}_S = \omega_q \hat{J}_z, 
\end{align}
where $\hat{J}_z = \frac{1}{2} \sum_{i=1}^L \hat{\sigma}_z^{(i)}$,
and $\hat{\sigma}_z^{(i)}$ is the $z$-component of the Pauli operator for the $i$-th qubit with a frequency $\omega_q$.
Then, for $\hat{J}_z$ and
an operator $\hat{J}^2$ that represents the total angular momentum,
we consider a subspace spanned by Dicke states
$\{ |L/2,M\rangle \}$,
which is a simultaneous
eigenstate of $\hat{J}^2$ and $\hat{J}_z$,
with eigenvalues $\frac{L}{2} \left( \frac{L}{2} + 1 \right)$ and $M$, respectively
($M = L/2, (L/2)-1, \ldots , -L/2$).
Throughout this Letter, we omit the label $L/2$ of the Dicke states,
i.e.,
$\ket{M} \equiv \ket{L/2, M}$.
Explicitly, the Dicke state $\ket{M}$ is a superposition state of all the computational states having $\frac{L}{2} + M $ excited states and $ \frac{L}{2} - M$ ground states with an equal coefficient, for an odd number $L$.

For an interaction between the $L$-qubit system and an environment
modeled by a single bath (i.e. $N_B=1$),
we consider an {\it $m$-body interaction}
($m=1, 2, \ldots , L$)~\footnote{See the subsection III~A of Supplementary Material for a detailed calculation.}.
In particular, we introduce the following system operator $\hat{A}$ for the interaction Hamiltonian $\hat{H}_{\text{int}} = \hat{A} \otimes \hat{B}$:
\begin{align}
    \hat{A} = \frac{g L}{ {}_L C_m } \biggl[ &\hat{\sigma}^{(1)}_x \otimes \cdots \otimes \hat{\sigma}^{(m)}_x \otimes \hat{I}^{(m+1)} \otimes \cdots \otimes \hat{I}^{(L)} \n
    & + \text{(all possible permutations)} \biggr],
\end{align}
where $g$ denotes a constant coupling strength,
and $\hat{\sigma}^{(i)}_x$ ($\hat{I}^{(i)}$) denotes the $x$-component of the Pauli (identity) operator for the $i$-th qubit.
Although $\hat{A}$ contains a total of
${}_L C_m = \frac{L !}{ (L-m) ! m! } $ terms,
we have $\| \hat{A} \| = \Theta (L)$ owing to the normalization by the prefactor;
thus, $\| \hat{H}_{\text{int}} \| = \Theta (L)$.

Under the Born and Markov approximations,
we derive a Redfield equation for the $L$-qubit system coupled with a white-noise Markovian bath.
Then, for an initial Dicke state
$\hat{\rho}_S (0) = \ket{L/2} \! \bra{L/2}$,
we obtain an instantaneous heat current $J (0)$ as follows:
\begin{align}
   J (0) = - \gamma_{\text{wn}}  \omega_q \frac{ L^2 m}{ {}_L C_m }  ,
\end{align}
where $\gamma_{\text{wn}}$ denotes a constant dissipation coefficient
for the white-noise environment.
For $m = L$,
the interaction Hamiltonian induces a direct transition from $\ket{L/2}$ (all-excited state) to $\ket{-L/2}$ (all-ground state),
and the absolute value of the heat current scales as $| J (0) | = \Theta (L^3)$.
Thus, the universal scaling bound $\Theta (L^3)$ becomes saturated.
More generally, for the case of $L-m = \Theta (L^0)$,
we can evaluate the scaling as $| J (0) | = \Theta ( L^{3+m-L} )$
by using an identity $ {}_L C_{m-1} = \frac{m}{L-m+1} {}_L C_{m} $.
Therefore, in our example, the construction of the interaction
for the universal bound $\Theta (L^3)$ to be saturated requires $L$-body interactions.


{\it{Bound 2.}}\textemdash
The previous example requires $L$-body interactions 
for the scaling bound to be saturated,
which might be difficult to realize with the existing technology.
Therefore, we consider more feasible cases here.
In practical situations, the transfer energy due to an interaction with an environment should be upper-bounded by a threshold value,
because the transitions should occur between not-too-distant energy levels of the system Hamiltonian.
In particular, we focus on a specific class of the system
by setting a constraint that any non-diagonal elements of the noise operators become zero in the system energy basis,
if the energy difference exceeds a certain value $\Delta E_k^{(a)}$ for channel $k$ of the $a$-th bath.
Mathematically, this reads as follows:
\begin{align}
    |E_i - E_j| > \Delta E^{(a)}_k  \Rightarrow \
    \bra{i} \hat{A}^{(a)}_k \ket{j} = 0 \ \
    \forall i,j,
    \label{eq:DeltaE}
\end{align}
where we define a spectral decomposition of the system Hamiltonian as
$\hat{H}_S = \sum_{i=1}^N E_i \ket{i} \! \bra{i}$,
and $N$ is the dimension of the Hilbert space of the system.
Note that the system is allowed to be degenerate, and a
similar assumption was made in Ref.~\cite{gyhm2022quantum}.
Under this condition, by using a relation that was also used in Ref.~\cite{gyhm2022quantum},
we rigorously derive another new bound on the absolute value of the heat current as follows~\footnote{See the subsection II~B of Supplementary Material for a detailed derivation of Bound 2.}:
\begin{align}
   | J_a (t) | \le 2 \sum_{k,l=1}^{ c_a } \Xi^{(a)}_{kl} \Delta E^{(a)}_k \| \hat{A}^{(a)}_k \| \| \hat{A}^{(a)}_l \|. 
    \label{eq:bound2}
\end{align}
Therefore, if we assume that $\Delta E_k^{(a)} = \Theta (L^0)$ ($\forall a, k$),
then $| J_a (t) |$ scales at most as $\Theta (L^2)$,
and this provides a different scaling bound from
that provided by Eq.~(\ref{eq:bound1}).
Moreover, owing to the constraint,
we can discuss more realistic cases for the bound to be saturated
without many-body interactions.


{\it{Bound 2 for superradiance and superabsorption.}}\textemdash
We discuss how our results can be applied to non-interacting qubits coupled with a common bath.
For $L$ qubits that are coupled with a single bath having a single channel
(i.e., $N_B=1$ and $c_a = 1$),
a system Hamiltonian $\hat{H}_{\text{SR}}$ and an interaction Hamiltonian $\hat{H}_{\text{int}}$ are respectively given as follows:
\begin{align}
    \hat{H}_{\text{SR}} &= \omega_q \hat{J}_z,
    \label{hsrone} \\
    \hat{H}_{\text{int}} &= 2 g \hat{J}_x \otimes \hat{B},
    \label{hsrtwo}
\end{align}
where $\hat{J}_x = \frac{1}{2} \sum_{i=1}^L \hat{\sigma}^{(i)}_x$,
$\omega_q$ is a qubit frequency,
$\hat{B}$ is a bath operator,
and $g$ is a coupling strength between the $L$-qubit system and the environment.
For this system,
the condition of Eq.~(\ref{eq:DeltaE}) is satisfied for $\Delta E = \omega_q$;
Consequently, $\Delta E = \Theta (L^0)$.
Then, using Eq.~(\ref{eq:bound2}), we obtain $ |J_a (t)|  \leq \Theta (L^2)$.

\begin{figure}
	\begin{center}
		\includegraphics[clip,width=7cm,bb=0 0 425 325]{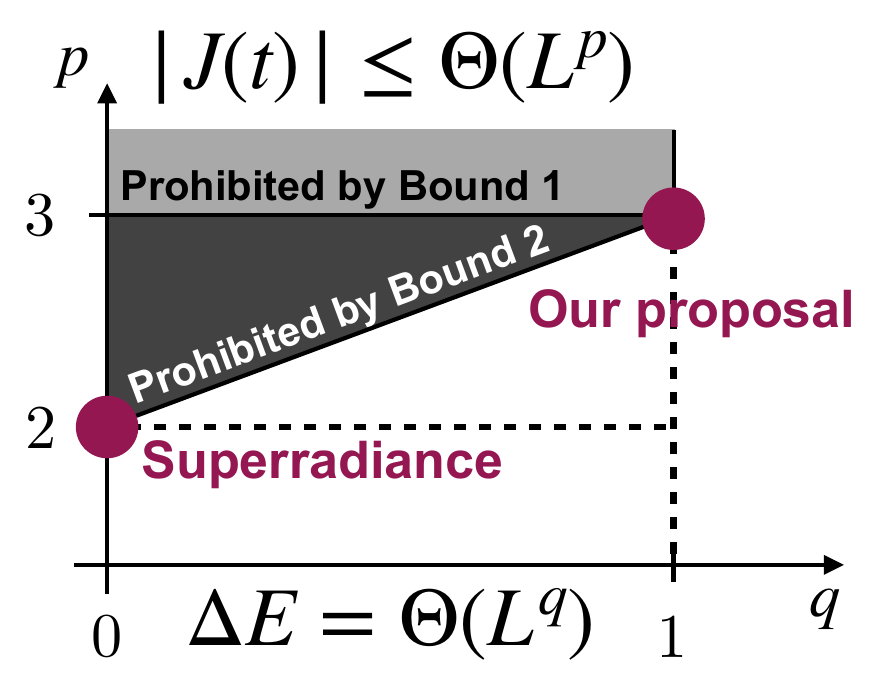}
		\caption{Scaling bounds on a heat current $J(t)$ flowing into an $L$-particle open system.
		The light gray (dark gray) region represents the prohibited scaling by Bound 1 (Bound 2).
		For simplicity, we consider the case of a single bath
		and introduce $\Delta E = \max_k \Delta E_k$.
		Superradiance causes Bound 2 to be saturated,
		whereas our proposal of using an $L$-body interaction
		causes Bound 1 to be saturated.
		Here, we assume the extensivity of the system Hamiltonian, i.e., $\| \hat{H}_S \| = \Theta (L)$.}
		\label{fig:bounds}
	\end{center}
\end{figure}

We compare our results with those of previous studies on
superradiance~\footnote{See the subsection III~B of Supplementary Material for a detailed calculation.}.
Suppose that $L$ is an odd number.
Under a Redfield equation derived for the system of superradiance with a zero temperature bath,
we obtain a heat current for an initial state $\ket{1/2}$ as follows:
\begin{align}
    J (0) = - \frac{1}{4} \gamma_0 \omega_q (L+1)^2,
\end{align}
where $\gamma_0$ denotes a constant dissipation coefficient.
This means that superradiance saturates the scaling bound $\Theta (L^2)$ obtained from Eq.~(\ref{eq:bound2}).
(see Fig.~\ref{fig:bounds} for a summary of our results).

Furthermore, we investigate the case of superabsorption,
which is interpreted as the reverse process of superradiance~\cite{higgins2014superabsorption}. 
The system Hamiltonian is expressed as follows:
\begin{align}
    \hat{H}_{\text{SA}} = \omega_q \hat{J}_z + \Omega \hat{J}_z^2.
\end{align}
A critical component of the superabsorption is the additionally introduced term $\Omega \hat{J}_z^2$.
Combined with an engineering of a spectral structure of the environment,
the system exhibits an energy absorption process from $\ket{ - 1/2 } $ to $ \ket{1/2} $ with a rate of $\Theta (L^2)$.
This is superabsorption.
However, the superabsorption does not cause the bound in
Eq.~(\ref{eq:bound2}) to be saturated in terms of scaling about $L$.
This is because the interaction induces a transition from $\ket{-L/2}$ to $\ket{-(L/2)+1}$,
and the energy difference between these two eigenstates is $\omega_q + (L-1) \Omega$.
To use the constraints in Eq.~(\ref{eq:DeltaE}),
we must set $\Delta E=\Theta (L)$ for this model.
Thus, Eq.~(\ref{eq:bound2}) provides an upper bound
that scales as $\Theta (L^3)$,
which cannot be saturated by superabsorption.

{\it{Bound on a steady-state heat current.}}\textemdash
Although, so far, we have
discussed the upper bounds on a heat current for a given initial state,
we can also derive an upper bound on the heat current in a steady state.
However, it is difficult to define a steady state for the Redfield equation 
due to the counter-rotating terms.
Thus, we adopt a rotating-wave approximation (RWA) for the Redfield equation,
and we obtain 
the following Gorini--Kossakowski--Sudarshan--Lindblad~(GKSL) master equation
$\frac{d \hat{\rho}_S (t)}{dt} = \mathcal{L} [ \hat{\rho}_S (t) ]$~\cite{breuer2002theory}:
\begin{align}
    \mathcal{L} [ \hat{\rho} ]
    = -i [ \hat{H}_S + \hat{H}_{\text{LS}} , \hat{\rho} ]
    + \sum_{a=1}^{N_B} \mathcal{D}_a^{\text{(G)}} [ \hat{\rho} ].
    \label{eq:GKSL}
\end{align}
Here, the Lamb shift term $\hat{H}_{\text{LS}}$ satisfies $[\hat{H}_S , \hat{H}_{\text{LS}} ] = 0$,
and the dissipator $\mathcal{D}_a^{\text{(G)}} $ is defined as follows:
\begin{align}
    \mathcal{D}_a^{\text{(G)}} [\hat{\rho} ] = \sum_{\omega} \sum_{k,l=1}^{c_a}
    \gamma_{kl, \omega }^{(a)}
    \left( \hat{A}_{l, \omega}^{(a)} \hat{\rho} \hat{A}_{k, \omega}^{(a) \dagger}
    - \frac{1}{2} \{ \hat{A}_{k, \omega}^{(a) \dagger} \hat{A}_{l, \omega}^{(a)} , \hat{\rho} \}\right),
\end{align}
where $\{ \hat{A}, \hat{B} \} = \hat{A} \hat{B} + \hat{B} \hat{A}$ is an anti-commutator.
Then, for the GKSL equation, we define a steady state $\hat{\rho}_{\text{ss}}$ for an initial state $\hat{\rho}_S (0)$
as follows:
\begin{align}
    \hat{\rho}_{\text{ss}} = \lim_{t \to \infty} e^{\mathcal{L} t} [ \hat{\rho}_S (0) ].
\end{align}
First, by using the commutation relationship
$[ \hat{H}_S , \hat{A}_{l,\omega}^{(a)}] = - \omega \hat{A}_{l,\omega}^{(a)}$~($\forall l, \omega, a$)
that is satisfied for a microscopically derived GKSL equation,
we show that
$[\hat{H}_S, \hat{\rho}_{\text{ss}} ] = 0 $
when $[ \hat{H}_S , \hat{\rho}_S (0)] = 0$.
Second, we show that, if $[\hat{H}_S , \hat{\rho} ] = 0$ is satisfied,
a heat current $ J_a^{\text{(G)}} ( \hat{\rho} ) = \text{Tr} ( \hat{H}_S \mathcal{D}_a^{\text{(G)}} [ \hat{\rho}] )$
calculated by the GKSL master equation (after the RWA) is the same as
that calculated by the Redfield equation (before the RWA)
$J_a (\hat{\rho}) = \text{Tr} ( \hat{H}_S \mathcal{D}_a [\hat{\rho}] )$.
Therefore, we can use the right-hand sides of
Eqs.~\eqref{eq:bound1}~and~\eqref{eq:bound2}
as the upper bounds on the steady-state heat current
$J_{\text{ss,}a} = \text{Tr} ( \hat{H}_S \mathcal{D}^{\text{(G)}}_a [\hat{\rho}_{\text{ss}}] )$,
which were originally derived for the Redfield equation.
Moreover, we find a specific system to show
a steady-state heat current of $| J_{\text{ss}}| = \Theta (L^3)$,
which saturates Bound~1 in terms of scaling.
Consequently, we can construct new quantum-enhanced thermodynamic devices
such as a quantum heat engine
whose power output $P$ scales as $P = \Theta (L^3)$
while its efficiency being fixed~\footnote{See the subsection IV~D of Supplementary Material for some details.}.


{\it{Conclusion.}}\textemdash
In this Letter, we discussed newly derived bounds on a heat current flowing into an open quantum system weakly coupled with an environment.
First, we derived a general scaling bound
(with the number of particles)
on the absolute value of the heat current by assuming
the extensivity of a system Hamiltonian and a system-environment interaction Hamiltonian.
In particular, we found that the best achievable scaling for the $L$-particle system is $\Theta (L^3)$ in a limit of large $L$.
However, for the scaling bound to be saturated in our example,
an $L$-body interaction is required, which may be difficult to realize with the existing technology.
Then, we derived another bound based on the constraint
that any non-diagonal elements of the noise operators become zero with respect to the system energy basis,
if the energy difference of these two bases exceeds a certain value $\Delta E$.
Based on this second bound, we showed that the absolute value of the heat current scales at most as $\Theta (L^2)$ for $\Delta E = \Theta (L^0)$,
and this bound is saturated for superradiance.
We first revealed the bounds that universally limit how fast a heat current generated by an open $L$-particle quantum system can scale with the number of particles.
Our results are applicable not only to an open quantum system involving an interaction between particles
but also to various types of the environment spectrum,
both of which are expected to improve the performance of quantum thermodynamic devices
such as quantum heat engines, quantum refrigerators, and quantum batteries.

We thank A. Yoshinaga for fruitful discussions.
This work was supported by MEXT's Leading Initiative for Excellent Young Researchers
and JST PRESTO (No.~JPMJPR1919), Japan.
Y.~T. acknowledges support from JSPS KAKENHI (No. 20H01827) and
JST's Moonshot R\&D (No.~JPMJMS2061).




\bibliographystyle{apsrev4-1}
\bibliography{QHEnogo_main}

\end{document}



\title{Supplementary Material: Universal Scaling Bounds on a Quantum Heat Current}


\author{Shunsuke Kamimura}
\email{s2130043@s.tsukuba.ac.jp}
\affiliation{Faculty of Pure and Applied Sciences, University of Tsukuba, Tsukuba 305-8571, Japan}
\affiliation{Research Center for Emerging Computing Technologies,  National  Institute  of  Advanced  Industrial  Science  and  Technology  (AIST),1-1-1  Umezono,  Tsukuba,  Ibaraki  305-8568,  Japan.}

\author{Kyo Yoshida}
\affiliation{Faculty of Pure and Applied Sciences, University of Tsukuba, Tsukuba 305-8571, Japan}

\author{Yasuhiro Tokura}
\email{tokura.yasuhiro.ft@u.tsukuba.ac.jp}
\affiliation{Faculty of Pure and Applied Sciences, University of Tsukuba, Tsukuba 305-8571, Japan}

\author{Yuichiro Matsuzaki}
\email{matsuzaki.yuichiro@aist.go.jp}
\affiliation{Research Center for Emerging Computing Technologies,  National  Institute  of  Advanced  Industrial  Science  and  Technology  (AIST),1-1-1  Umezono,  Tsukuba,  Ibaraki  305-8568,  Japan.}

\maketitle

In Supplementary Material, we provide some detailed discussion
about the present work.
In Sec.~I, we set up a model of an open quantum system
and derive a system equation of motion,
which is referred to as a Redfield equation in the literature,
based on Born and Markov approximations.
In Sec.~II, we derive the universal scaling bounds on a heat current.
In Sec.~III, we analyze two examples
to saturate the bound of our inequalities in terms of scaling.
In Sec.~IV, we
discuss how we can bound a steady-state heat current for a Gorini-Kossakowski-Sudarshan-Lindblad (GKSL) master equation system, and
present
new quantum enhanced thermodynamic devices,
a quantum heat engine and a quantum battery,
which are inspired by the example introduced in Sec.~III.
Finally, in Sec.~V, we address an experimentally feasible demonstration of our newly derived bound (Bound~1)
that is saturated by a many-body interaction between particles induced by their environment.

\section{Open Quantum System: Redfield equation}
\label{sec:openqsys}

In this section, we derive a Redfield equation according to Ref.~\cite{breuer2002theory}.

\subsection{Hamiltonian}

The total system under consideration
is composed of a quantum system surrounded by an environment that is modeled by $N_B$ heat baths.
The total Hamiltonian is given as follows:
\begin{align}
	\hat{H}_{\text{tot}} = \hat{H}_S + \hat{H}_B + \hat{H}_{\text{int}}. 
\end{align}
We explicitly represent the form of $\hat{H}_{\text{int}}$
by sets of Hermitian operators $\{ \hat{A}_{k,a} \}$ and $\{ \hat{B}_{k,a} \}$ as
\begin{align}
	\hat{H}_{\text{int}} &= \sum_{a=1}^{N_B} \hat{H}_{\text{int} , a}, &
	\hat{H}_{\text{int}, a} &= \sum_{k=1}^{c_a} \hat{A}_{k,a} \otimes \hat{B}_{k,a}.
	\label{eq:explicit_form_interaction}
\end{align}
We also express $\hat{H}_B$ as a summation of local Hamiltonians
$\hat{H}_{B,a}$ acting on the $a$-th bath: 
\begin{align}
	\hat{H}_B = \sum_{a=1}^{N_B} \hat{H}_{B,a}.
\end{align}
Furthermore, we assume that the
$a$-th bath is in a thermal equilibrium state with an inverse temperature $\beta_a$.
In this case, the quantum state $\hat{\rho}_B$ of the environment is expressed as a product state of these thermal equilibrium states:
\begin{align}
	\hat{\rho}_B &= \bigotimes_{a=1}^{N_B} \hat{\rho}_{B, a}, &
	\hat{\rho}_{B, a} &= \frac{1}{Z_{B,a} } e^{- \beta_a \hat{H}_{B,a} } ,
	\ \ \ Z_{B,a} = \text{Tr}_a [e^{- \beta_a \hat{H}_{B,a} } ] .
\end{align}
For the bath operator $\hat{B}_{k,a} $ in the interaction Hamiltonian
in Eq.~(\ref{eq:explicit_form_interaction}),
let us introduce the following condition:
\begin{align}
	\text{Tr}_a [ \hat{B}_{k,a} \hat{\rho}_{B,a} ]  = 0 \ \ \ \forall k, a.
\end{align}
Without loss of generality,
we assume that this condition holds for the following reason:
even if our original interaction Hamiltonian does not satisfy this condition for $\hat{B}_{k,a} $,
we can recover it 
by modifying the system Hamiltonian $\hat{H}_S$ and the operator $\hat{B}_{k,a}$ as 
\begin{align}
	\hat{H}_S &\mapsto \hat{H}_S + \sum_{a=1}^{N_B}  \sum_{ k=1 }^{c_a} \hat{A}_{k,a} \text{Tr}_a [ \hat{B}_{k,a} \hat{\rho}_{B,a} ] , &
	\hat{B}_{k,a} &\mapsto \hat{B}_{k,a} - \text{Tr}_a [ \hat{B}_{k,a}  \hat{\rho}_{B,a} ] .
\end{align}

\subsection{Redfield equation based on Born and Markov approximations}

Under the above setup,
in the interaction picture with respect to $\hat{H}_0 = \hat{H}_S + \hat{H}_B$,
(which is represented by the superscript~${}^{(I)}$)
we have dynamics of a quantum state of the total system $\hat{\rho}_{\text{tot}} (t)$ governed by the following von-Neumann equation ($\hbar = 1$):
\begin{align}
	\frac{d}{dt} \hat{\rho}_{\text{tot}}^{(I)} (t) = -i [ \hat{H}_{\text{int}}^{(I)} (t), \hat{\rho}_{\text{tot}}^{(I)} (t) ],
\end{align}
where
\begin{align}
	\hat{H}_{\text{int}}^{(I)} (t) &= e^{ i \hat{H}_0 t } \hat{H}_{\text{int}} e^{ - i \hat{H}_0 t }, &
	\hat{ \rho }_{\text{tot}}^{(I)} (t) = e^{ i \hat{H}_0 t } \hat{ \rho }_{\text{tot}} (t) e^{ - i \hat{H}_0 t }.
\end{align}
By arranging the von-Neumann equation and tracing out the environmental degrees of freedom, we obtain the following exact equation for the reduced density operator $\hat{\rho}^{(I)}_{S} (t) = \text{Tr}_B [ \hat{\rho}^{(I)}_{\text{tot}} (t)  ]$ of the system:
\begin{align}
	\frac{d }{ dt} \hat{\rho}^{(I)}_S (t)
	&= - \int_0^t ds \text{Tr}_B \left(  \left[ \hat{H}^{(I)}_{\text{int}} (t) , [ \hat{H}^{(I)}_{\text{int}} (s) , \hat{\rho}^{(I)}_{\text{tot}} (s) ] \right] \right),
	\label{eq:dynamics_redop}
\end{align}
with the initial condition $\hat{\rho}_{\text{tot}} (0) = \hat{\rho}_S (0) \otimes \hat{\rho}_B$.

In the integral of the right-hand side of Eq.~(\ref{eq:dynamics_redop}), 
by adopting the Born approximation,
which is expressed as $\hat{\rho}^{(I)}_{\text{tot}} (s) \simeq \hat{\rho}^{(I)}_S (s) \otimes \hat{\rho}_B$,
we obtain
\begin{align}
	\frac{d}{dt} \hat{\rho}^{(I)}_S (t) 
	= - \int_0^t ds \text{Tr}_B \left(  \left[ \hat{H}^{(I)}_{\text{int}} (t) , [ \hat{H}^{(I)}_{\text{int}} (s) , \hat{\rho}^{(I)}_S (s) \otimes \hat{\rho}_B ] \right] \right).
	\label{eq:dynamics_Born}
\end{align}

Next, we apply the Markov approximation,
in which we execute the replacement
$\hat{\rho}_S^{(I)} (s) \mapsto \hat{\rho}_S^{(I)} (t)$,
change the variable $s \mapsto t-s$,
and render the upper limit $t$ of the integral infinity.
Then, we obtain the so-called Redfield equation:
\begin{align}
	\frac{d}{dt} \hat{\rho}^{(I)}_S (t) 
	= - \int_0^{\infty} ds \text{Tr}_B \left(  \left[ \hat{H}^{(I)}_{\text{int}} (t) , [ \hat{H}^{(I)}_{\text{int}} (t-s) , \hat{\rho}^{(I)}_S ( t ) \otimes \hat{\rho}_B ] \right] \right).
	\label{eq:MarkovEq_Breuer3118}
\end{align}


Recalling the explicit definition of the interaction Hamiltonian $\hat{H}_{\text{int}}$ in Eq.~(\ref{eq:explicit_form_interaction})
and executing the partial trace $\text{Tr}_B$ for the bath degrees of freedom,
we have another form of the Redfield equation as
\begin{align}
	\frac{d}{dt} \hat{\rho}^{(I)}_S (t) 
	= \int_0^{\infty} ds  \sum_{a,b=1}^{N_B} \sum_{k=1}^{c_a} \sum_{l=1}^{c_b} C_{klab} (t-s) 
	\left[ - \hat{A}_{k,a}^{(I)} (t) \hat{A}_{l,b}^{(I)} (s) \hat{\rho}_S^{(I)} (t)
	+ \hat{A}_{l,b}^{(I)} (s)  \hat{\rho}_S^{(I)} (t) \hat{A}_{k,a}^{(I)} (t) \right]
	+ \text{h.c.} ,
	\label{eq:Redfield_withC}
\end{align}
where ``h.c." represents an Hermitian conjugate.
Here, we define the correlation function
$C_{klab} (t-s) = \text{Tr}_B \left[ \hat{B}_{k,a}^{(I)} (t) \hat{B}_{l,b}^{(I)} (s) \hat{\rho}_B \right]$,
which is determined only by the time difference $t-s$,
because
\begin{align}
	C_{klab} (t-s) = 
	\text{Tr}_B \left[ \hat{B}_{k,a}^{(I)} (t) \hat{B}_{l,b}^{(I)} (s) \hat{\rho}_B \right]
	=  \text{Tr}_B \left[ \hat{B}_{k,a}^{(I)} (t-s) \hat{B}_{l,b}^{(I)} (0) \hat{\rho}_B \right]
	= \text{Tr}_B \left[ \hat{B}_{k,a}^{(I)} (t-s) \hat{B}_{l,b} \hat{\rho}_B \right], 
\end{align}
by virtue of the property $ [ \hat{H}_B , \hat{\rho}_B ] = 0 $.
Moreover, by using the condition $\text{Tr}_a [ \hat{B}_{k,a} \hat{\rho}_{B,a} ]  = 0$
($\forall k, a$),
we can find that $C_{klab} (t-s) = 0$ for $a \neq b$.
Using this, we can simplify the summation $\sum_{b=1}^{N_B} $ and obtain
\begin{align}
	\frac{d}{dt} \hat{\rho}^{(I)}_S (t) 
	= \int_0^{\infty} ds  \sum_{a=1}^{N_B} \sum_{k,l=1}^{c_a} C_{klaa} (t-s) 
	\left[ - \hat{A}_{k,a}^{(I)} (t) \hat{A}_{l,a}^{(I)} (s) \hat{\rho}_S^{(I)} (t)
	+ \hat{A}_{l,a}^{(I)} (s)  \hat{\rho}_S^{(I)} (t) \hat{A}_{k,a}^{(I)} (t) \right]
	+ \text{h.c.} .
	\label{eq:MarkovEq}
\end{align}
Going back to the original Schr\"odinger picture,
for the quantum state
$\hat{\rho}_S (t) = e^{- i \hat{H}_S t } \hat{\rho}_S^{(I)} (t) e^{i \hat{H}_S t}$ of the system,
the Redfield equation can be rewritten as
\begin{align}
	\frac{d}{dt} \hat{\rho}_S (t) =  -i [ \hat{H}_S , \hat{\rho}_S (t) ]
	+  \int_0^{\infty} d s  \sum_{a=1}^{N_B} \sum_{k,l=1}^{c_a}  \left[ C_{klaa} (s) 
	\left\{ - \hat{A}_{k,a} \hat{A}_{l,a}^{(I)} (- s) \hat{\rho}_S (t)
	+ \hat{A}_{l,a}^{(I)} ( -s)  \hat{\rho}_S (t) \hat{A}_{k,a} \right\} 
	+ \text{h.c.} \right].
	\label{eq:Redfield_Schr}
\end{align}

\section{Universal scaling bounds on a quantum heat current}
\label{sec:universal_bound}

In this section, based on the Redfield equation
considered in the previous section,
we derive
Bound 1 and Bound 2 in the main text.
Both of these theorems are the main results of this work.

\subsection{Derivation of Bound 1}



We define a heat current $J_a (t)$ from the $a$-th bath to the system as
\begin{align}
	J_a (t) \ceq \text{Tr} \left( \hat{H}_S \mathcal{D}_a [ \hat{\rho}_S (t) ] \right),
\end{align}
where the dissipator $\mathcal{D}_a $ acting on the system density operator is defined
as
\begin{align}
	\mathcal{D}_a [ \hat{\rho} ]
	\ceq  \int_0^{ \infty } d s \sum_{k,l=1}^{c_a}  C_{klaa} (s) 
	\left[ - \hat{A}_{k,a} \hat{A}_{l,a}^{(I)} (- s) \hat{\rho}
	+ \hat{A}_{l,a}^{(I)} ( - s) \hat{\rho} \hat{A}_{k,a} \right]
	+ \text{h.c.} .
\end{align}
This owes to the Redfield equation in Eq.~(\ref{eq:Redfield_Schr}).
Regarding this heat current, we can derive Bound 1 as follows:
\begin{align}
	&\left| J_a (t)  \right| \n
	&= \left| \int_0^{ \infty } d s  \sum_{k,l=1}^{c_a} C_{klaa} (s) 
	\text{Tr}  \left[ \hat{H}_S \left( - \hat{A}_{k,a} \hat{A}_{l,a}^{(I)} (- s) \hat{\rho}_S (t)
	+ \hat{A}_{l,a}^{(I)} ( -s) \hat{\rho}_S (t) \hat{A}_{k,a} \right)  \right]
	+ \text{c.c.} \right| \n
	&\le 2 \left|
	\int_0^{ \infty } d s  \sum_{k,l=1}^{c_a}
	C_{klaa} (s)
	\text{Tr}  \left[ 
	\hat{H}_S \left( 
	- \hat{A}_{k,a} \hat{A}_{l,a}^{(I)} (- s) \hat{\rho}_S (t)
	+ \hat{A}_{l,a}^{(I)} ( -s) \hat{\rho}_S (t) \hat{A}_{k,a} 
	\right) 
	\right]
	\right| 
	&& (\because | z + \text{c.c.} | \le 2 | z| ) \n
	&\le 2 \int_0^{ \infty } d s  \sum_{k,l=1}^{c_a}
	\left| C_{klaa} (s) \right|
	\left|
	\text{Tr}  \left[ 
	\hat{H}_S \left( 
	- \hat{A}_{k,a} \hat{A}_{l,a}^{(I)} (- s) \hat{\rho}_S (t)
	+ \hat{A}_{l,a}^{(I)} ( -s) \hat{\rho}_S (t) \hat{A}_{k,a} 
	\right) 
	\right]
	\right| \n
	&\le 2 \int_0^{\infty} d s \sum_{k,l=1}^{c_a} \left| C_{klaa} (s) \right| \left\| [ \hat{A}_{k,a} , \hat{H}_S ] \hat{A}_{l,a}^{(I)} (- s) \right\| \ && (\because | \text{Tr} [ \hat{A} \hat{\rho} ] | \le \| \hat{A} \| ) \n
	& \le 2 \int_0^{\infty} d s \sum_{k,l=1}^{c_a} \left| C_{klaa} (s) \right|
	\left\| [ \hat{A}_{k,a} , \hat{H}_S ] \right\| 
	\left\| \hat{A}_{l,a}^{(I)} (- s) \right\| 
	&& (\because \| \hat{A} \hat{B} \| \le \| \hat{A} \| \| \hat{B} \| )\n
	&= 2 \int_0^{\infty} d s \sum_{k,l=1}^{c_a} \left| C_{klaa} (s) \right|
    \left\| [ \hat{A}_{k,a} , \hat{H}_S ] \right\| 
	\| \hat{A}_{l,a} \| 
	&& \left( \because \| \hat{U} \hat{A} \hat{U}^{\dagger} \| = \| \hat{A} \| \right)
	\label{ineq_inBound1} \\
	&\le 4 \sum_{k,l=1}^{c_a} \left( \int_0^{\infty} d s  \left| C_{klaa} (s) \right| \right)
	\| \hat{A}_{k,a} \| \| \hat{H}_S \| \| \hat{A}_{l,a} \| ,
	&& ( \because \| \hat{A} \hat{B} - \hat{B} \hat{A} \| \le 2 \| \hat{A} \| \| \hat{B} \| )
\end{align}
where we used the properties of the operator norm $\| \bullet \|$
defined as
$\| \hat{O} \| = \sup_{\ket{\psi} } \frac{ \| \hat{O} \ket{\psi} \| }{ \| \ket{\psi} \| }$.
Introducing a coefficient $\Xi_{kl}^{(a)}$ as
\begin{align}
    \Xi_{kl}^{(a)} \ceq \int_0^{\infty} d s  \left| C_{klaa} (s) \right|,
\end{align}
then we obtain Bound 1 presented as Eq.~(9) in the main text:
\begin{align}
    | J_a (t) | \le 4 \| \hat{H}_S \|  \sum_{k,l=1}^{c_a} \Xi_{kl}^{(a)} \| \hat{A}_{k,a} \| \| \hat{A}_{l,a} \| .
\end{align}


\subsection{Derivation of Bound 2}
To derive Bound 2, we go back to the inequality (\ref{ineq_inBound1})
that appears during the derivation of Bound 1:
\begin{align}
    \left| J_a (t)  \right| \le 2 \sum_{k,l=1}^{c_a} \Xi_{kl}^{(a)} \left\| [ \hat{A}_{k,a} , \hat{H}_S ] \right\| 
	\| \hat{A}_{l,a} \| .
	\label{ineq2_inBound1}
\end{align}
According to the notation of the main text, we introduce a spectral decomposition of the system Hamiltonian as
\begin{align}
    \hat{H}_S = \sum_{i=1}^N E_i
   \ket{i} \! \bra{i}. 
\end{align}
As done in the main text, we impose the following condition on $\hat{A}_{k,a}$ and $\hat{H}_S$:
\begin{align}
    |E_i - E_j| > \Delta E^{(a)}_k  \Rightarrow
    \bra{i} \hat{A}_{k,a} \ket{j} = 0 \ \
    \forall i, j .
    \label{eq:DeltaE}
\end{align}
For an Hermitian operator $\hat{A}_{k,a}$ satisfying the constraint in Eq.~(\ref{eq:DeltaE}),
we can derive the following nontrivial inequality after some involved calculations:
\begin{align}
    \left\| [ \hat{A}_{k,a} , \hat{H}_S ] \right\| \le \Delta E_{k}^{(a)} \| \hat{A}_{k,a} \|.
    \label{eq:inequality_DeltaE}
\end{align}
This inequality is formulated and derived as {\bf Theorem 1} in Supplemental Material of Ref.~\cite{gyhm2022quantum}.
Combining this inequality with Eq.~(\ref{ineq2_inBound1}),
we obtain
\begin{align}
     \left| J_a (t)  \right| \le 2 \sum_{k,l=1}^{c_a} \Xi_{kl}^{(a)} \Delta E_{k}^{(a)} \| \hat{A}_{k,a} \|
	\| \hat{A}_{l,a} \| .
\end{align}
This is Bound 2, which is presented as Eq.~(14) in the main text.
\section{Examples of Redfield equations in the main text}
\label{sec:Appl_main_text}

\subsection{Heat current for a many-body interaction}
\label{subsec:mbody_heat}
In this subsection,
we discuss an example to saturate Bound 1.
Specifically, we focus on an $L$-qubit system
involved in a many-body interaction mediated by
a zero-temperature white-noise environment,
and calculate a heat current generated by the environment.
We consider the following system Hamiltonian $\hat{H}_S$
composed of identical $L$ qubits with a frequency $\omega_{q}$:
\begin{align}
    \hat{H}_S = \omega_{q} \hat{J}_z
    = \frac{\omega_q}{2} \sum_{j=1}^L \hat{\sigma}^{(j)}_z.
\end{align}
For the system-environment interaction Hamiltonian
$\hat{H}_{\text{int}} = \hat{A} \otimes \hat{B}$,
we assume that this induces an $m$-body interaction between the qubits.
Mathematically, we introduce a set $\mathcal{S}_m$
that contains all the combinations of $m$ qubits out of 
the $L$ qubits:
\begin{align}
    \mathcal{S}_m = \left\{ (i_1, i_2, \ldots , i_m ) \left| \ i_j \in \{ 1, 2, \ldots , L \}, \
    i_1 < i_2 < \cdots < i_m \right. \right\} .
\end{align}
Then, we define the system noise operator $\hat{A}$ as
\begin{align}
    \hat{A} = \frac{g L}{ {}_L C_m }
    \sum_{ \bm{i} \in \mathcal{S}_m }
    \left[ \hat{\sigma}_x^{(i_1)} \otimes \hat{\sigma}_x^{(i_2)} \otimes \cdots \otimes \hat{\sigma}_x^{(i_m)} \otimes 
    \hat{I}^{( \overline{\bm{i}} )} \right],
\end{align}
where $\hat{I}^{( \overline{\bm{i}} )}$ represents
an identity operator acting onto the $(L-m)$-qubit degrees of freedom
that are not contained in $\bm{i} \in \mathcal{S}_m$.
Although the operator $\hat{A}$ contains ${}_L C_m$ terms in total,
the interaction Hamiltonian is extensive,
i.e. $\| \hat{H}_{\text{int}} \| = gL = \Theta (L)$.
This is because the state $\ket{+ + \ldots +}$ is an eigenvector of each operator $\hat{\sigma}_x^{(i_1)} \otimes \hat{\sigma}_x^{(i_2)} \otimes \cdots \otimes \hat{\sigma}_x^{(i_m)} \otimes \hat{I}^{( \overline{\bm{i}} )} $ with the maximum eigenvalue 1,
and thus this state is also an eigenvector of $\hat{A}$ with the maximum eigenvalue $gL$,
where $\ket{+}$ denotes an eigenvector of $\hat{\sigma}_x$ with an eigenvalue $1$ for each qubit.
The interaction Hamiltonian is written as
\begin{align}
    \hat{H}_{\text{int}}
    = \hat{A} \otimes \hat{B}
    = \frac{g L}{ {}_L C_m }
    \sum_{ \bm{i} \in \mathcal{S}_m }
    \left[ \hat{\sigma}_x^{(i_1)} \otimes \hat{\sigma}_x^{(i_2)} \otimes \cdots \otimes \hat{\sigma}_x^{(i_m)} \otimes 
    \hat{I}^{( \overline{\bm{i}} )} \right] \otimes \hat{B}.
\end{align}
Under this setup, we can derive a Redfield equation.
In the interaction picture, the system noise operator $\hat{A}$
is transformed as
\begin{align}
    \hat{A}^{(I)} (-s) &= e^{ - i \hat{H}_S s} \hat{A} e^{ + i \hat{H}_S s} \n
    &= \frac{g L}{ {}_L C_m } \sum_{ \bm{i} \in \mathcal{S}_m }
    e^{ - \frac{ i \omega_q s}{2} \sum_{j=1}^L \hat{\sigma}_z^{(j)} } 
    \left[ \hat{\sigma}_x^{(i_1)} \otimes \hat{\sigma}_x^{(i_2)} \otimes \cdots \otimes \hat{\sigma}_x^{(i_m)} \otimes 
    \hat{I}^{( \overline{\bm{i}} )} \right]
    e^{ + \frac{ i \omega_q s}{2} \sum_{j=1}^L \hat{\sigma}_z^{(j)} } \n
    &= \frac{g L}{ {}_L C_m }
    \sum_{ \bm{i} \in \mathcal{S}_m }
    \left[ \hat{\sigma}_+^{(i_1)} e^{-i \omega_q s} + \hat{\sigma}_-^{(i_1)} e^{+i \omega_q s} \right] \otimes \cdots \otimes
    \left[ \hat{\sigma}_+^{(i_m)} e^{-i \omega_q s} + \hat{\sigma}_-^{(i_m)} e^{+i \omega_q s} \right] \otimes
    \hat{I}^{( \overline{\bm{i}} )}.
\end{align}
Then, we calculate a heat current
$ J (t) = \text{Tr} \left( \hat{H}_S \mathcal{D} [ \hat{\rho}_S (t) ] \right)$
under the following Redfield equation:
\begin{align}
    \frac{d}{dt} \hat{\rho}_S (t) &= 
    -i  [ \hat{H}_S, \hat{\rho}_S (t) ] 
    + \mathcal{D}[ \hat{\rho}_S (t) ],  \\
    & \text{where} \ \mathcal{D} [\hat{\rho}] = \int_0^{\infty} ds \ C (s) \left[ \hat{A}^{(I)} (-s) \hat{\rho} \hat{A} - \hat{A} \hat{A}^{(I)} (-s) \hat{\rho} \right] + \text{h.c.} .
\end{align}
In particular, for an all-excited Dicke state in the initial time,
i.e. $\hat{\rho}_S (0)
= \left| \frac{L}{2} \right> \! \left\langle \frac{L}{2} \right|$,
the heat current $J (0)$ is written as
\begin{align}
    J (0) = \int_0^{\infty} ds \ C(s)
    \left[ \text{Tr} \left( \hat{H}_S \hat{A}^{(I)} (-s) \left| \frac{L}{2} \right> \! \left\langle \frac{L}{2} \right| \hat{A} - \hat{H}_S \hat{A} \hat{A}^{(I)} (-s) \left| \frac{L}{2} \right> \! \left\langle \frac{L}{2} \right| \right) \right] + \text{c.c.} .
\end{align}
On the other hand, remembering the definition of the Dicke state,
the Dicke state $\left| \frac{L}{2} - m \right\rangle$
is a superposition state with an equal coefficient
for all the computational states having $m$ qubits in
the ground state and $(L-m)$ ones in the excited state.
Equivalently,
\begin{align}
    \left| \frac{L}{2} - m \right\rangle &= \frac{1}{ \sqrt{ {}_L C_m } } \sum_{ \bm{i} \in \mathcal{S}_m }
    \hat{\sigma}_-^{(i_1)} \otimes \hat{\sigma}_-^{(i_2)} \otimes \ldots \hat{\sigma}_-^{(i_m)}
    \otimes \hat{I}^{( \overline{\bm{i}} )}
    \left| \frac{L}{2} \right\rangle ,
    &
    m = 0, 1, \ldots , L-1, L.
\end{align}
Based on this, we have the following two identities:
\begin{align}
    \hat{A} \left| \frac{L}{2} \right\rangle
    &= \frac{ gL }{ {}_L C_m } \sum_{ \bm{i} \in \mathcal{S}_m }
    \hat{\sigma}_x^{(i_1)} \otimes \hat{\sigma}_x^{(i_2)} \otimes \ldots \hat{\sigma}_x^{(i_m)}
    \otimes \hat{I}^{( \overline{\bm{i}} )}
    \left| \frac{L}{2} \right\rangle \n
    &= \frac{ gL }{ {}_L C_m } \sum_{ \bm{i} \in \mathcal{S}_m }
    \hat{\sigma}_-^{(i_1)} \otimes \hat{\sigma}_-^{(i_2)} \otimes \ldots \hat{\sigma}_-^{(i_m)}
    \otimes \hat{I}^{( \overline{\bm{i}} )}
    \left| \frac{L}{2} \right\rangle \n
    &= \frac{gL}{ {}_L C_m } \sqrt{ {}_L C_m } \left| \frac{L}{2} - m \right\rangle \n
    &= \frac{gL}{ \sqrt{ {}_L C_m } } \left| \frac{L}{2} - m \right\rangle,
\end{align}
and
\begin{align}
    \hat{A}^{(I)} (-s) \left| \frac{L}{2} \right\rangle
    &= \frac{ gL }{ {}_L C_m } \sum_{ \bm{i} \in \mathcal{S}_m }
    \left[ \hat{\sigma}_+^{(i_1)} e^{-i \omega_q s} + \hat{\sigma}_-^{(i_1)} e^{+i \omega_q s} \right] \otimes \cdots \otimes
    \left[ \hat{\sigma}_+^{(i_m)} e^{-i \omega_q s} + \hat{\sigma}_-^{(i_m)} e^{+i \omega_q s} \right] \otimes
    \hat{I}^{( \overline{\bm{i}} )}
    \left| \frac{L}{2} \right\rangle \n
    &= \frac{ gL }{ {}_L C_m } \sum_{ \bm{i} \in \mathcal{S}_m }
    e^{ i m \omega_q s }
    \hat{\sigma}_-^{(i_1)} \otimes \hat{\sigma}_-^{(i_2)} \otimes \ldots \hat{\sigma}_-^{(i_m)}
    \otimes \hat{I}^{( \overline{\bm{i}} )}
    \left| \frac{L}{2} \right\rangle \n
    &= \frac{gL}{ \sqrt{ {}_L C_m } }
    e^{ i m \omega_q s } \left| \frac{L}{2} - m \right\rangle.
\end{align}
Using these identities, we can evaluate the heat current as
\begin{align}
    J (0) &= \int_0^{\infty} ds \ C (s)
    \left[ \text{Tr} \left( \hat{H}_S \hat{A}^{(I)} (-s) \left| \frac{L}{2} \right\rangle \! \left\langle \frac{L}{2} \right| \hat{A} - \hat{H}_S \hat{A} \hat{A}^{(I)} (-s) \left| \frac{L}{2} \right\rangle \! \left\langle \frac{L}{2} \right|
    \right) \right]
    + \text{c.c.} \n
    &= \frac{ g^2 L^2 }{ {}_L C_m } \int_0^{\infty} ds \
    e^{i m \omega_q s} C (s)
    \left[ \left\langle \frac{L}{2} - m \right| \hat{H}_S \left| \frac{L}{2} - m \right\rangle 
    - E_{\frac{L}{2} }
    \right] + \text{c.c.} \n
    &= \frac{ g^2 L^2 }{ {}_L C_m } \int_0^{\infty} ds \
    e^{i m \omega_q s} C (s) \left( E_{\frac{L}{2} - m} - E_{\frac{L}{2} } \right) + \text{c.c.} \n
    &= - g^2 \omega_q \frac{ L^2 m }{ {}_L C_m } \int_{-\infty}^{\infty} ds \
    e^{i m \omega_q s} C (s),
\end{align}
where we used the property $C^*(s) = C(-s)$,
and $E_M = M \omega_q$ is an energy of the Dicke state $\ket{M}$
($M = -(L/2), -(L/2)+1, \ldots , (L/2)-1, (L/2)$).
Here, for simplicity,
we assume that the Fourier transform of the correlation function $C (s)$
(i.e. spectral function)
for the white-noise environment is constant about the frequency:
\begin{align}
  g^2 \int_{-\infty}^{\infty} ds \
    e^{i m \omega_q s} C (s)
    =  \gamma_{\text{wn}} = \text{const.} ,
\end{align}
which means that the dissipation rate is determined regardless of the frequency of transition that the environment induces.
Then, we obtain
\begin{align}
   J (0) = - \gamma_{\text{wn}} \omega_q \frac{ L^2 m }{ {}_L C_m }.
\end{align}
This result is the same as
Eq.~(12) in the main text.
Moreover, we obtain $J(0)=\Theta (L^3)$ for $m=L$,
which saturates Bound 1 in terms of scaling.

\subsection{Heat current for a superradiance}
\label{subsec:superradiance}
In this subsection, we detail the calculation of
a heat current for a system exhibiting superradiance,
and this saturates Bound~2.
We assume the following system and interaction Hamiltonians $\hat{H}_{\text{SR}}$
and
$\hat{H}_{\text{int}}$:
\begin{align}
    \hat{H}_{\text{SR}} &= \omega_{q} \hat{J}_z ,
    &
    \hat{H}_{\text{int}} &= 2 g \hat{J}_x \otimes \hat{B}
    = g \left[ \sum_{i=1}^L \hat{\sigma}_x^{(i)} \right] \otimes \hat{B},
\end{align}
which are the same as those introduced in the main text.
Under this setup, we obtain the following Redfield equation
(See Eq.~(\ref{eq:Redfield_Schr}) for the generic expression):
\begin{align}
    \frac{d}{dt} \hat{\rho}_S (t) = -i [ \hat{H}_{\text{SR}} , \hat{\rho}_S (t) ]
    + \int_0^{\infty} ds \ C (s) \left[ \left\{ \hat{A}^{(I)} (-s) \hat{\rho}_S (t) \hat{A} - \hat{A} \hat{A}^{(I)} (-s) \hat{\rho}_S (t) \right\}
    + \text{h.c.} \right],
\end{align}
where, for the present system, we have
\begin{align}
    \hat{A} &= 2 g \hat{J}_x ,
    &
    C (s) = \text{Tr}_B [ \hat{B}^{(I)} (s) \hat{B} \hat{\rho}_{B} ].
\end{align}
Let us especially focus on an instantaneous density operator
\begin{align}
    \hat{\rho}_S (t) = \sum_{m = - L/2}^{L/2} p_m (t) \ket{m} \! \bra{m}
\end{align}
that is diagonal with respect to the Dicke states $\{ \ket{m} \equiv \ket{\frac{L}{2}, m} \}$ with the maximum total angular momentum.
The operator $\hat{J}_x = \frac{1}{2} ( \hat{J}_+ + \hat{J}_- )$
can be simplified based on how the operator $\hat{J}_{\pm}$ acts onto the Dicke states as
\begin{align}
    \hat{J}_{+} &= \sum_{m = -L/2}^{L/2}
     C_{m,+} \ket{m+1} \! \bra{m} ,\\
     \hat{J}_{-} &= \sum_{m = -L/2}^{L/2}
     C_{m,-} \ket{m-1} \! \bra{m},
\end{align}
where the coefficient $C_{m, \pm}$ is defined as
\begin{align}
    C_{m, \pm} = \sqrt{ \left( \frac{L}{2} \mp m \right) \left( \frac{L}{2} \pm m +1 \right) }.
\end{align}
Then, by using the explicit forms of the operators 
\begin{align}
    \hat{A} &= 2 g \hat{J}_x = g \left[ \hat{J}_+ + \hat{J}_- \right], \\
    \hat{A}^{(I)} (-s) &= g \left[ e^{ -i \omega_{q} s } \hat{J}_+ +  e^{ +i \omega_{q} s } \hat{J}_- \right],
\end{align}
we can obtain the following identities:
\begin{align}
    \hat{A} \ket{l} &= g \left[ C_{l, +} \ket{l+1} + C_{l, -} \ket{l-1} \right], \\
    \hat{A}^{(I)} (-s) \ket{m} &= g \left[ e^{- i \omega_{q} s} C_{m, +} \ket{m+1} + e^{+ i \omega_{q} s} C_{m, -} \ket{m-1} \right].
\end{align}
Thus, we can calculate a heat current $J (t)$ for the state
$ \hat{\rho}_S (t) = \sum_{m = - L/2}^{L/2} p_m (t) \ket{m} \! \bra{m}$ as
\begin{align}
    J (t) &= \text{Tr} \left[ \hat{H}_{\text{SR}} \hat{\rho}_S (t) \right] \n
    &= \int_0^{\infty} ds \ C (s)
    \left(
    \text{Tr} \left[ \hat{H}_{\text{SR}} \hat{A}^{(I)} (-s) \hat{\rho}_S (t) \hat{A} \right]
    -\text{Tr} \left[ \hat{H}_{\text{SR}} \hat{A} \hat{A}^{(I)} (-s) \hat{\rho}_S (t) \right]
    \right) + \text{c.c.} \n
    &= \sum_{l,m= -L/2}^{L/2} E_l p_m (t) \int_0^{\infty} ds \ C(s) \bra{l} \hat{A}^{(I)} (-s) \ket{m} \! \bra{m} \hat{A} \ket{l} \n
    & \ \ \ \ \ - \sum_{l,m= -L/2}^{L/2} E_l p_m (t) \int_0^{\infty} ds \ C(s) \bra{l} \hat{A} \hat{A}^{(I)} (-s) \ket{m} \! \braket{m |l} + \text{c.c.} \n
    &= g^2 \sum_{m= -L/2}^{L/2} p_m \left[ E_{m-1} C_{m-1, +} C_{m, -} \int_0^{\infty} ds \ C(s) e^{ i \omega_{q} s}
    + E_{m+1} C_{m+1, -} C_{m, +} \int_0^{\infty} ds \ C(s) e^{ - i \omega_{q} s} \right] \n
    & \ \ \ \ \ + g^2 \sum_{m= -L/2}^{L/2} E_{m} p_m \left[ \left( C_{m, +} \right)^2 \int_0^{\infty} ds \ C(s) e^{ -i \omega_{q} s}
    + \left( C_{m, -} \right)^2 \int_0^{\infty} ds \ C(s) e^{ i \omega_{q} s} \right] \n
    & \ \ \ \ \ \ \ \ \ \ + \text{c.c.} \n
    &=  \sum_{m=-L/2}^{L/2} p_m \left[ E_{m-1} C_{m-1, +} C_{m, -} \gamma (\omega_{q}) +E_{m+1} C_{m+1, -} C_{m, +} \gamma ( - \omega_{q})
    - E_m \left\{ \left( C_{m, +} \right)^2 \gamma  ( - \omega_{q}) + \left( C_{m, -} \right)^2 \gamma  ( \omega_{q}) \right\} \right],
\end{align}
where we introduce a spectral function $\gamma ( \omega )$ as
\begin{align}
    \gamma (\omega)
    &= g^2 \int_{-\infty}^{\infty} ds \ C (s) e^{ i \omega s}.
\end{align}
This spectral function $\gamma (\omega)$ satisfies the following detailed-balance condition for the heat bath with an inverse temperature $\beta$:
\begin{align}
    \frac{\gamma (\omega) }{\gamma (-\omega) } = e^{\beta \omega}.
\end{align}
In particular, we focus on a case in which the number of qubits $L$ is odd.
In a scenario of the superradiance,
one often assumes an initial Dicke state
$\hat{\rho}_S (0) = \ket{1/2} \! \bra{1/2}$
(i.e. $p_m (0) = \delta_{m, 1/2}$),
and a zero temperature bath (i.e. $\beta \to \infty$) for simplicity.
Then, by neglecting the excitation rate $\gamma (-\omega_{q})$
that is exponentially smaller than the relaxation rate $\gamma_0 = \gamma (\omega_{q})$,
a heat current for the initial state $J (0)$ is simplified as
\begin{align}
    J (0) &= \gamma_0 \left[ E_{-1/2} C_{-1/2, +} C_{1/2, -}
    - E_{1/2} \left( C_{1/2, -} \right)^2 \right]
    = \gamma_0 \left( E_{-1/2} - E_{1/2} \right) \times \frac{1}{4} (L+1)^2 \n
    &= - \frac{1}{4} \gamma_0 \omega_{q} (L+1)^2,
\end{align}
where we use the definition $E_m = m \omega_{q}$.
This result is the same as
Eq.~(17) in the main text.
Because the noise operator $\hat{A} = 2 g \hat{J}_x$ only induces transitions
between $\ket{m}$ and $\ket{m \pm 1}$,
we can take
$\Delta E = |E_{m} - E_{m \pm 1} | = \omega_q = \Theta (L^0)$.
Thus,
the obtained heat current
$|J(0) | =\Theta (L^2)$
saturates Bound 2 in terms of scaling.

\section{GKSL master equation, steady-state heat current, and quantum thermodynamic devices}
\label{sec:GKSL}

In this section, we consider the so-called
Gorini--Kossakowski--Sudarshan--Lindblad (GKSL) master equation,
which we derive based on Ref.~\cite{breuer2002theory}, with an additional approximation called a rotating-wave approximation (RWA).
For this system, we have a steady state of the open quantum system,
and discuss a steady-state heat current.
In particular, we can bound the steady-state heat current by utilizing the bounds
derived for the Redfield equation system
under some assumptions.
Inspired by the example (that saturates Bound 1) in Sec.~III,
we find models of quantum heat engine and quantum battery
that exhibit quantum enhanced performances,
and we will explain
them in this section.


\subsection{GKSL master equation based on a rotating-wave approximation}
Suppose that the system Hamiltonian $\hat{H}_S$,
whose spectral decomposition is
$\hat{H}_S = \sum^N_{i=1} E_i \ket{i} \! \bra{i}$, is
decomposed as
\begin{align}
	\hat{H}_S = \sum_{\epsilon} \epsilon \hat{\Pi} ({\epsilon}),
\end{align}
where $\epsilon$ represents an energy of the system Hamiltonian,
and
$\hat{\Pi} ( \epsilon ) = \sum_{i; E_i = \epsilon } \ket{i} \! \bra{i}  $
is a projector onto the corresponding energy eigenspace.
Let us define an operator $\hat{A}_{k,a}$ as
\begin{align}
	\hat{A}_{k,a, \omega} &\ceq \sum_{\epsilon, \epsilon' ; \epsilon - \epsilon' = - \omega} \hat{\Pi} ({\epsilon}) \hat{A}_{k,a} \hat{\Pi} ({\epsilon'}),
\end{align}
which turns out to be a Lindblad operator in the GKSL master equation, and satisfies the commutation relation
\begin{align}
    [ \hat{H}_S , \hat{A}_{k,a,\omega} ] = - \omega \hat{A}_{k,a, \omega} ,
    \ \ \ \forall k,a, \omega.
    \label{eq:commutation}
\end{align}
Then, by using these operators in the Redfield equation,
we have
\begin{align}
	\frac{d}{dt} \hat{\rho}^{(I)}_S (t) &= \sum_{a=1}^{N_B} \sum_{k, l=1}^{c_a}  \sum_{\omega, \omega'}
	\Gamma^{(a)}_{kl} (\omega) e^{ i (\omega' - \omega) t  }
	\left\{ \hat{A}_{l,a, \omega} \hat{\rho}^{(I)}_S (t)  \hat{A}_{k, a, \omega'}^{\dagger} - \hat{A}_{k, a,\omega'}^{\dagger} \hat{A}_{l, a, \omega} \hat{\rho}^{(I)}_S (t) \right\}
	+ \text{h.c.} .
	\label{eq:MarkovEq_beforeRWA1}
\end{align}
Here, we define a one-sided Fourier transform $\Gamma^{(a)}_{kl} (\omega) $ of the previously defined correlation function as 
\begin{align}
	\Gamma^{(a)}_{kl} (\omega) &\ceq \int_0^{\infty} ds e^{i \omega s } C_{klaa} (s).
\end{align}
At last, we adopt the rotating-wave (secular) approximation.
In Eq.~(\ref{eq:MarkovEq_beforeRWA1}), we have oscillating factors $e^{ i ( \omega' - \omega ) t}$ in front of the system density operator $\hat{\rho}^{(I)}_S (t)$.
The RWA is validated under the condition where the frequency of oscillation $ |\omega' - \omega| $ for $\omega \neq \omega'$ is much larger than 
the typical rate of the system dynamics, which is roughly estimated by $\| d \hat{\rho}^{(I)}_S (t)/ dt \|$,
or more concretely, by the system-environment coupling strength.
Then, we can ignore the terms for $\omega' \neq \omega$ because the contributions from such fast-oscillating terms are cancelled out after the integration about time $t$.
Eventually, we end up with the following GKSL master equation in the interaction picture:
\begin{align}
	\frac{d}{dt} \hat{\rho}^{(I)}_S (t) 
	= -i [ \hat{H}_{\text{LS}} ,\hat{\rho}^{(I)}_S (t)   ] + \sum_{a=1}^{N_B} \mathcal{D}^{(a)}_{\text{G}} [\hat{\rho}^{(I)}_S (t) ] ,
\end{align}
where we define the dissipator $\mathcal{D}_{\text{G}}^{(a)}$ for the $a$-th bath as
\begin{align}
	\mathcal{D}_{\text{G}}^{(a)} [ \hat{\rho} ] \ceq \sum_{\omega} \sum_{k,l=1}^{c_a} \gamma^{(a)}_{kl} (\omega)  \left( \hat{A}_{l, a, \omega}   \hat{\rho}   \hat{A}_{k, a, \omega}^{\dagger} - \frac{1}{2} \{ \hat{A}_{k, a, \omega}^{\dagger} \hat{A}_{l, a, \omega},    \hat{\rho}  \} \right).
	\label{eq:dissipatorDa_exp}
\end{align}
Also, we define as
\begin{align}
	\gamma^{(a)}_{kl} (\omega) &= \Gamma^{(a)}_{kl} (\omega) + \Gamma^{(a)*}_{ lk } (\omega), \\
	S^{(a)}_{kl} (\omega) &= \frac{1}{2i} \left( \Gamma^{(a)}_{kl} (\omega) - \Gamma^{(a) *}_{ lk } (\omega) \right) , \\
	\hat{H}_{\text{LS}} &= \sum_{a=1}^{N_B} \sum_{\omega} \sum_{k,l=1}^{c_a} S^{(a)}_{kl} (\omega) \hat{A}_{k, a, \omega }^{\dagger} \hat{A}_{l, a, \omega },
\end{align}
where $\hat{H}_{\text{LS}}$ describes the Lamb shift.
In the original Schr\"odinger picture, the system dynamics are given by
\begin{align}
	\frac{d}{dt} \hat{\rho}_S(t) = -i [ \hat{H}_S + \hat{H}_{\text{LS}} , \hat{\rho}_S (t) ] 
	+ \sum_{a=1}^{N_B} \sum_{\omega } \sum_{k,l=1}^{c_a} \gamma^{(a)}_{kl} (\omega) \left( \hat{A}_{l, a, \omega} \hat{\rho}_S (t)  \hat{A}_{k, a, \omega}^{\dagger} - \frac{1}{2} \{ \hat{A}_{k, a, \omega}^{\dagger} \hat{A}_{l, a, \omega},  \hat{\rho}_S (t) \} \right) , 
	\label{eq:GKSL_full-app}
\end{align}
which is guaranteed by the relation $[ \hat{H}_S, \hat{H}_{\text{LS}} ] = 0$ and $e^{ i \hat{H}_S t } \hat{A}_{k, a, \omega} e^{ - i \hat{H}_S t } = e^{ - i \omega t } \hat{A}_{k, a, \omega}$ $\forall k$.
Here, we introduced the anti-commutator $\{ \hat{A}, \hat{B} \} = \hat{A} \hat{B} + \hat{B} \hat{A}$.

Along with the convention of quantum thermodynamics,
we ignore the contribution from Lamb shift $\hat{H}_{\text{LS}}$.
This is reasonable because the Lamb shift $\hat{H}_{\text{LS}}$ does not contribute to a heat current $J (t) = \text{Tr} [ \hat{H}_S (d \hat{\rho}_S (t) / dt ) ]$ at all,
due to the property $[ \hat{H}_S, \hat{H}_{\text{LS}} ] = 0$.

\subsection{First and second law of thermodynamics}

We define a net heat flow $J (t)$ from the environment to the system as
a time derivative of the internal energy $E_{\text{sys}} (t) = \text{Tr} [ \hat{H}_S \hat{\rho}_S (t) ]$
measured by the system Hamiltonian $\hat{H}_S$.
Namely,
\begin{align}
	J (t)
	&\ceq \frac{d E_{\text{sys}} (t) }{dt}
	= \text{Tr} \left( \hat{H}_S \frac{d \hat{\rho}_S (t) }{dt} \right)
	= \sum_{a=1}^{N_B} \text{Tr} \left( \hat{H}_S \mathcal{D}_{\text{G}}^{(a)} [ \hat{\rho}_S (t) ] \right) .
\end{align}
In the last equality, we use the GKSL master equation~(\ref{eq:GKSL_full-app})
and the relation $\text{Tr} \left( \hat{H}_S [ \hat{H}_S + \hat{H}_{\text{LS}} , \hat{\rho}_S (t) ] \right) = 0$.
Defining a heat current $J_{\text{G}}^{(a)} (t)$ from the $a$-th bath to the system as
$J_{\text{G}}^{(a)} (t) = \text{Tr} \left( \hat{H}_S \mathcal{D}_{\text{G}}^{(a)} [ \hat{\rho}_S (t) ] \right) $,
we have
\begin{align}
	J (t) = \sum_{a=1}^{N_B} J_{\text{G}}^{(a)} (t).
	\label{eq:first_law}
\end{align}
This is the first law of quantum thermodynamics.
In particular, for a steady state $\hat{\rho}_{\text{s.s.}} $ of the GKSL equation, satisfying $\mathcal{L} [ \hat{\rho}_{\text{s.s.}} ] = 0$,
we have $J (t) = 0$, and then the first law is expressed as
\begin{align}
	\sum_{a=1}^{N_B} J_{\text{G}}^{(a)} (t) = 0.
	\label{eq:first_law_forss}
\end{align}


Then, we describe what the second law of thermodynamics claims and how this can be applicable for concrete problems.
Here, we accept the second law without any proof, whereas a detailed proof for a slightly different GKSL equation from Eq.~(\ref{eq:GKSL_full-app}) is presented in Supplementary Material of Ref.~\cite{tajima2021superconducting}.

We define an entropy production rate $\dot{\sigma} (t)$, which quantifies a degree of dissipation of the system, as
\begin{align}
	\dot{\sigma} (t) = \frac{d S (t) }{dt} - \sum_{a=1}^{N_B} \beta_a J_{\text{G}}^{(a)} (t).
\end{align}
Here, $d S(t)/dt$ is the time derivative of the von-Neumann entropy
$S (t) = \text{Tr} [\hat{\rho}_S (t) \log \hat{\rho}_S (t)]$ of the system.
For this entropy production rate $\dot{\sigma} (t)$, the second law of thermodynamics claims the following inequality:
\begin{align}
	\dot{\sigma} (t) \ge 0.
\end{align}



In a situation where we have three heat baths $a = \text{H}$, C and W ($=$ work),
and the quantum state of the system is a steady state of the GKSL master equation,
then $\dot{S} (t) = 0$ and the second law of thermodynamics is expressed as
$ \beta_{\text{H}} J_{\text{H}} (t) + \beta_{\text{C}} J_{\text{C}} (t) + \beta_{\text{W}} J_{\text{W}} (t)  \ge 0$.

Suppose that, in the regime of a heat engine,
the three heat currents take the sign as
\begin{align}
	J_{\text{H}} (t) &\ge 0,
	& J_{\text{C}} (t) &\le 0, 
	& J_{\text{W}} (t) &\le 0.
\end{align}
We further assume that the temperature of the W (work) reservoir is much higher than the others; $\beta_{\text{W}} \ll \beta_{\text{H}}, \beta_{\text{C}}$.
Then, the second law is approximately described as $\beta_{\text{H}} J_{\text{H}} (t) + \beta_{\text{C}} J_{\text{C}} (t) \ge 0$.
Combining these conditions with the first law $J_{\text{H}} (t) + J_{\text{C}} (t) + J_{\text{W}} (t)  = 0$,
we obtain the Carnot bound $\eta_{\text{Carnot}}$ for the efficiency $\eta (t)$ of the heat engine:
\begin{align}
	\eta (t) \ceq \frac{ - J_{\text{W}} (t) }{ J_{\text{H}} (t) } = 1 + \frac{ J_{\text{C}} (t)  }{ J_{\text{H}} (t) } \le 1 - \frac{ \beta_{\text{H}} }{  \beta_{\text{C}} } = \eta_{\text{Carnot}} .
\end{align}

Similarly, let us consider the regime of a refrigerator;
\begin{align}
	J_{\text{H}} (t) &\le 0, & J_{\text{C}} (t) &\ge 0, & \ J_{\text{W}} (t) &\ge 0.
\end{align}
Then, the Carnot bound of the COP (coefficient of performance) is derived as
\begin{align}
	\epsilon_{\text{COP}} (t) \ceq \frac{ J_{\text{C}} (t) }{ J_{\text{W}} (t) } = \frac{ J_{\text{C}} (t) }{ -J_{\text{H}} (t) -J_{\text{C}} (t) }
	= \frac{ 1 }{ -1 - \frac{J_{\text{H}} (t)}{ J_{\text{C}} (t)} } \le \frac{ 1 }{ -1 + \frac{\beta_{\text{C}} }{ \beta_{\text{H}} } } = \frac{ \beta_{\text{H}} }{ \beta_{\text{C}} - \beta_{\text{H}} } = \epsilon_{\text{Carnot}}.
\end{align}

\subsection{Scaling bound on a steady-state heat current for GKSL master equation system}

\subsubsection{Relationship between a heat current for the Redfield equation and that for the GKSL master equation}

Here, we discuss how we can bound a steady-state heat current in a GKSL master equation system.
First, recalling Eq.~(\ref{eq:MarkovEq_beforeRWA1}),
we write down the Redfield equation in the Sch\"odinger picture as follows:
\begin{align}
    \frac{d}{dt} \hat{\rho}_S (t) = - i [ \hat{H}_S , \hat{\rho}_S (t) ]
    + \sum_{a=1}^{N_B} \mathcal{D}_{\text{R}}^{(a)} [ \hat{\rho}_S (t) ],
\end{align}
where the dissipator $\mathcal{D}_{\text{R}}^{(a)}$ for the Redfield equation is defined as
\begin{align}
    \mathcal{D}_{\text{R}}^{(a)} [ \hat{\rho} ]
    \ceq \sum_{k, l=1}^{c_a}  \sum_{\omega, \omega'}
	\Gamma^{(a)}_{kl} (\omega) \left( \hat{A}_{l,a, \omega} \hat{\rho} \hat{A}_{k, a, \omega'}^{\dagger} - \hat{A}_{k, a,\omega'}^{\dagger} \hat{A}_{l, a, \omega} \hat{\rho} \right)
	+ \text{h.c.}
    \label{eq:Redfield_dissipator}
\end{align}
Applying the rotating-wave approximation,
we restrict the summation only for $\omega = \omega'$,
and obtain the following GKSL master equation:
\begin{align}
    \frac{d}{dt} \hat{\rho}_S (t)
    = \mathcal{L} [\hat{\rho}_S (t)]
    \ceq - i [ \hat{H}_S , \hat{\rho}_S (t) ]
    + \sum_{a=1}^{N_B} \mathcal{D}_{\text{G}}^{(a)} [ \hat{\rho}_S (t) ],
    \label{eq:GKSL_lindbladian}
\end{align}
where the dissipator $\mathcal{D}_{\text{G}}^{(a)}$ for the GKSL master equation is defined as
\begin{align}
    \mathcal{D}_{\text{G}}^{(a)} [ \hat{\rho} ]
    \ceq \sum_{k, l=1}^{c_a}  \sum_{\omega}
	\Gamma^{(a)}_{kl} (\omega) \left( \hat{A}_{l,a, \omega} \hat{\rho} \hat{A}_{k, a, \omega}^{\dagger} - \hat{A}_{k, a,\omega}^{\dagger} \hat{A}_{l, a, \omega} \hat{\rho} \right)
	+ \text{h.c.} 
    \label{eq:GKSL_dissipator}
\end{align}
Note that the dissipator in Eq.~(\ref{eq:GKSL_dissipator}) contains the information about the Lamb shift,
and this expression is equivalent to Eq.~(\ref{eq:GKSL_full-app}).
Here, we use this expression to make comparison between the Redfield equation and the GKSL equation simpler.

For the dissipator of the Redfield equation in Eq.~(\ref{eq:Redfield_dissipator}),
we define a heat current flowing into the system of the density operator $\hat{\rho}$ from bath $a$ as
\begin{align}
    J_{\text{R}}^{(a)} (\hat{\rho}) \ceq \text{Tr} \left[ \hat{H}_S \mathcal{D}_{\text{R}}^{(a)} [\hat{\rho}] \right]
    =
    \sum_{k, l=1}^{c_a}  \sum_{\omega, \omega'}
	\Gamma^{(a)}_{kl} (\omega)
    \left( 
    \text{Tr} \left[ \hat{H}_S \hat{A}_{l,a, \omega} \hat{\rho} \hat{A}_{k, a, \omega'}^{\dagger} \right]
    - \text{Tr} \left[ \hat{H}_S \hat{A}_{k, a,\omega'}^{\dagger} \hat{A}_{l, a, \omega} \hat{\rho} \right]
    \right)
	+ \text{c.c.}
\end{align}
Similarly, we define a heat current for the GKSL master equation as
\begin{align}
    J_{\text{G}}^{(a)} (\hat{\rho}) \ceq \text{Tr} \left[ \hat{H}_S \mathcal{D}_{\text{G}}^{(a)} [\hat{\rho}] \right]
    =
    \sum_{k, l=1}^{c_a}  \sum_{\omega}
	\Gamma^{(a)}_{kl} (\omega)
    \left( 
    \text{Tr} \left[ \hat{H}_S \hat{A}_{l,a, \omega} \hat{\rho} \hat{A}_{k, a, \omega}^{\dagger} \right]
    - \text{Tr} \left[ \hat{H}_S \hat{A}_{k, a,\omega}^{\dagger} \hat{A}_{l, a, \omega} \hat{\rho} \right]
    \right)
	+ \text{c.c.}
\end{align}
Then, we can show the following theorem between these quantities:
\begin{align}
    [ \hat{H}_S, \hat{\rho} ] = 0
    \ \ \ \Rightarrow \ \ \
    J_{\text{R}}^{(a)} (\hat{\rho}) = J_{\text{G}}^{(a)} (\hat{\rho}).
    \label{eq:thm1_JR_JG}
\end{align}

Before we proceed the proof in Eq.~(\ref{eq:thm1_JR_JG}),
let us prove the following lemma:
\begin{align}
    [ \hat{H}_S, \hat{\rho} ] = 0
    \ \ \ \Rightarrow \ \ \
    \text{Tr} \left[ \hat{A}_{l,a, \omega} \hat{\rho} \hat{A}_{k, a, \omega'}^{\dagger} \right] = 0
    \ \ \ \text{for} \ \omega \neq \omega'.
    \label{eq:prop_omega_neq}
\end{align}
To prove this lemma, we first remember the explicit expressions of the
Lindblad operators $\hat{A}_{l,a, \omega}$ and $\hat{A}_{k,a,\omega'}^{\dagger}$ as follows:
\begin{align}
	\hat{A}_{l,a,\omega} &\ceq \sum_{ \substack{ e, e'; \\ e - e' = \omega } } \hat{\Pi} (e') \hat{A}_{l,a} \hat{\Pi} (e)
	= \sum_{ \substack{ e, e'; \\ e-e' = \omega } } \sum_{ \substack{ i; \ E_i = e \\ i'; \ E_{i'} = e' } } \ket{ i' } \! \bra{i' } \hat{A}_{l,a} \ket{i} \! \bra{i}  , \\
	\hat{A}_{k,a,\omega'}^{\dagger} &\ceq \sum_{ \substack{ \epsilon, \epsilon'; \\ \epsilon - \epsilon' = \omega' } } \hat{\Pi} (\epsilon) \hat{A}_{k,a} \hat{\Pi} (\epsilon')
	= \sum_{ \substack{ \epsilon, \epsilon'; \\ \epsilon-\epsilon' = \omega ' } } \sum_{ \substack{ j ; \ E_j = \epsilon \\ j'; \ E_{j'} = \epsilon' } } \ket{ j } \! \bra{j } \hat{A}_{k,a} \ket{j' } \! \bra{j' }.
\end{align}
Here, the spectral decomposition for the system Hamiltonian was given as
$\hat{H}_S = \sum_{i=1}^N E_i \ket{i} \! \bra{i}$, and
the projector $\hat{\Pi} (e)$ onto the energy eigenspace was defined as
\begin{align}
    \hat{\Pi} (e) \ceq \sum_{i; \ E_i = e } \ket{i} \! \bra{i}.
\end{align}
Then, on the one hand, the factor
$\text{Tr} \left[ \hat{A}_{l,a, \omega} \hat{\rho} \hat{A}_{k, a, \omega'}^{\dagger} \right]$
in Eq.~(\ref{eq:prop_omega_neq}) can be expressed as
\begin{align}
    \text{Tr} \left[ \hat{A}_{l,a, \omega} \hat{\rho} \hat{A}_{k, a, \omega'}^{\dagger} \right]
    &=
	\sum_{ \substack{ e, e'; \\ e-e' = \omega } }
	\sum_{ \substack{ i; \ E_i = e \\ i'; \ E_{i'} = e' } }
	\sum_{ \substack{ \epsilon, \epsilon'; \\ \epsilon-\epsilon' = \omega ' } }
	\sum_{ \substack{ j ; \ E_j = \epsilon \\ j'; \ E_{j'} = \epsilon' } }
	\text{Tr} \left[  
	 \ket{ i' } \! \bra{ i' } \hat{A}_{l,a} \ket{i} \! \bra{i} 
	 \hat{\rho}
	\ket{ j } \! \bra{ j } \hat{A}_{k,a} \ket{ j' } \! \bra{ j' }
	\right] \n
	&= 
	\sum_{ \substack{ e, e'; \\ e-e' = \omega } }
	\sum_{ \substack{ i; \ E_i = e \\ i'; \ E_{i'} = e' } }
	\sum_{ \substack{ \epsilon, \epsilon'; \\ \epsilon-\epsilon' = \omega ' } }
	\sum_{ \substack{ j ; \ E_j = \epsilon \\ j'; \ E_{j'} = \epsilon' } }
	\delta_{i' j'} \bra{ i } \hat{\rho} \ket{ j }
	\bra{ i' } \hat{A}_{l,a} \ket{ i } \!
	 \bra{ j } \hat{A}_{k,a} \ket{ j' }.
  \label{eq:factor_expression}
\end{align}
On the other hand, for the density operator $\hat{\rho}$,
we have the following property:
\begin{align}
    [ \hat{H}_S , \hat{\rho} ] = 0 
	\ \ \ 
	&\Leftrightarrow 
	\ \ \ 
	\bra{i} [ \hat{H}_S , \hat{\rho} ] \ket{j} = 0 \ \ \ \forall i,j \n
	&\Leftrightarrow 
	\ \ \ 
	(E_i - E_j ) \bra{i} \hat{\rho} \ket{j} = 0 \ \ \ \forall i,j \n
	&\Leftrightarrow 
	\ \ \ 
	``\bra{i} \hat{\rho} \ket{j} = 0 \ \ \ \forall i, j \ \text{s.t.} \ E_i \neq E_j  ".
\end{align}
In other words, a matrix element $\bra{i} {\hat{\rho}} \ket{j}$
can take a non-zero value only for two energy eigenbases $\ket{i}$ and $\ket{j}$ satisfying $E_i = E_j$,
when the density operator commutes with the system
Hamiltonian.
However, due to the conditions of summation
$E_i - E_{i'} = \omega$,
$E_j - E_{j'} = \omega'$
and $i' = j'$,
we can show that $E_i \neq E_{j}$ when $\omega \neq \omega'$,
and then $\bra{i} \hat{\rho} \ket{j} = 0$ follows.
Therefore, all the terms in Eq.~(\ref{eq:factor_expression}) should be
zero when $\omega \neq \omega'$,
and we now have proved the property in Eq.~(\ref{eq:prop_omega_neq}).

From now on, we give the proof of 
the theorem in Eq.~(\ref{eq:thm1_JR_JG}).
First, we make the expression of $J_{\text{R}}^{(a)} (\hat{\rho})$ simpler,
by successively using the commutation relation
$[ \hat{H}_S, \hat{A}_{l,a,\omega} ] = - \omega \hat{A}_{l,a,\omega}$,
the assumption $[\hat{H}_S, \hat{\rho}] = 0$,
and the trace cyclicity, as
\begin{align}
    J_{\text{R}}^{(a)} (\hat{\rho}) &=
    \sum_{k, l=1}^{c_a}  \sum_{\omega, \omega'}
    \Gamma^{(a)}_{kl} (\omega)
    \left( 
    \text{Tr} \left[ ( \hat{A}_{l,a, \omega} \hat{H}_S - \omega \hat{A}_{l,a, \omega} ) \hat{\rho} \hat{A}_{k, a, \omega'}^{\dagger} \right]
    - \text{Tr} \left[ \hat{H}_S \hat{A}_{k, a,\omega'}^{\dagger} \hat{A}_{l, a, \omega} \hat{\rho} \right]
    \right)
	+ \text{c.c.} \n
     &=
    \sum_{k, l=1}^{c_a}  \sum_{\omega, \omega'}
    \Gamma^{(a)}_{kl} (\omega)
    \left( 
    - \omega \text{Tr} \left[ \hat{A}_{l,a, \omega} \hat{\rho} \hat{A}_{k, a, \omega'}^{\dagger} \right]
    + \text{Tr} \left[ \hat{A}_{l,a, \omega} \hat{H}_S \hat{\rho} \hat{A}_{k, a, \omega'}^{\dagger} \right]
    - \text{Tr} \left[ \hat{H}_S \hat{A}_{k, a,\omega'}^{\dagger} \hat{A}_{l, a, \omega} \hat{\rho} \right]
    \right)
	+ \text{c.c.} \n
    &= 
    - \sum_{k, l=1}^{c_a}  \sum_{\omega, \omega'}
    \omega \Gamma^{(a)}_{kl} (\omega) \text{Tr} \left[ \hat{A}_{l,a, \omega} \hat{\rho} \hat{A}_{k, a, \omega'}^{\dagger} \right]
    + \sum_{k, l=1}^{c_a}  \sum_{\omega, \omega'}
    \Gamma^{(a)}_{kl} (\omega)
    \left( 
    \text{Tr} \left[ \hat{A}_{l,a, \omega} \hat{\rho} \hat{H}_S \hat{A}_{k, a, \omega'}^{\dagger} \right]
    - \text{Tr} \left[ \hat{H}_S \hat{A}_{k, a,\omega'}^{\dagger} \hat{A}_{l, a, \omega} \hat{\rho} \right]
    \right)
	+ \text{c.c.} \n
    &= - \sum_{k, l=1}^{c_a}  \sum_{\omega, \omega'}
    \omega \Gamma^{(a)}_{kl} (\omega) \text{Tr} \left[ \hat{A}_{l,a, \omega} \hat{\rho} \hat{A}_{k, a, \omega'}^{\dagger} \right]
    + \text{c.c.}
\end{align}
Thus, we can also express the heat current for the GKSL equation as
\begin{align}
    J_{\text{G}}^{(a)} (\hat{\rho}) 
    &= - \sum_{k, l=1}^{c_a}  \sum_{\omega}
    \omega \Gamma^{(a)}_{kl} (\omega) \text{Tr} \left[ \hat{A}_{l,a, \omega} \hat{\rho} \hat{A}_{k, a, \omega}^{\dagger} \right]
    + \text{c.c.}
\end{align}
By using the lemma in Eq. \eqref{eq:prop_omega_neq}, we show that the heat current for the Redfield equation is the same as that for the GKSL equation when $ [ \hat{H}_S, \hat{\rho} ] = 0$.

\subsubsection{Bound on a steady-state heat current}

Here, we discuss upper bounds on a heat current for a steady state
of the GKSL master equation.
First, we prove the following property for the Lindbladian $\mathcal{L}$
that is defined in Eq.~(\ref{eq:GKSL_dissipator}):
\begin{align}
    [ \hat{H}_S , \hat{\rho} ] = 0
    \ \ \ \Rightarrow \ \ \
    \left[ \hat{H}_S , \mathcal{L} [ \hat{\rho} ] \right] = 0.
    \label{eq:Lindbladian_commute}
\end{align}
In other words, when the operator $\hat{\rho}$ commutes with the system Hamiltonian,
then another operator $\mathcal{L} [ \hat{\rho} ]$ that the Lindbladian acts on
also commutes with the system Hamiltonian.

To prove the property in Eq.~(\ref{eq:Lindbladian_commute}),
we assume $[ \hat{H}_S, \hat{\rho} ] = 0$.
Then, by definition in Eq.~(\ref{eq:GKSL_dissipator}),
\begin{align}
    \left[\hat{H}_S, \mathcal{L} [ \hat{\rho} ] \right]
    &= -i \left[ \hat{H}_S, [ \hat{H}_S , \hat{\rho} ] \right]
    + \left\{ \sum_{a=1}^{N_B} \sum_{k, l=1}^{c_a} \sum_{\omega}
	\Gamma^{(a)}_{kl} (\omega) \left(
    [ \hat{H}_S, \hat{A}_{l,a, \omega} \hat{\rho} \hat{A}_{k, a, \omega}^{\dagger} ]
    - [ \hat{H}_S , \hat{A}_{k, a,\omega}^{\dagger} \hat{A}_{l, a, \omega} \hat{\rho} ] \right)
	- \text{h.c.} 
    \right\}
\end{align}
The first term $\left[ \hat{H}_S, [ \hat{H}_S, \hat{\rho} ] \right]$ in the right-hand side is trivially zero when $[ \hat{H}_S, \hat{\rho} ] = 0$.
For the other terms, by using the properties
$[ \hat{H}_S, \hat{A}_{l,a,\omega} ] = -\omega \hat{A}_{l,a,\omega}$
and
$[ \hat{H}_S, \hat{A}_{k,a,\omega}^{\dagger} ] = + \omega \hat{A}_{k,a,\omega}^{\dagger}$
as well,
we obtain
\begin{align}
    \hat{H}_S \hat{A}_{l,a, \omega} \hat{\rho} \hat{A}_{k, a, \omega}^{\dagger}
    &= \hat{A}_{l,a, \omega} \hat{H}_S \hat{\rho} \hat{A}_{k, a, \omega}^{\dagger} - \omega \hat{A}_{l,a, \omega} \hat{\rho} \hat{A}_{k, a, \omega}^{\dagger} \n
    &= \hat{A}_{l,a, \omega} \hat{\rho} \hat{H}_S \hat{A}_{k, a, \omega}^{\dagger} - \omega \hat{A}_{l,a, \omega} \hat{\rho} \hat{A}_{k, a, \omega}^{\dagger} \n
    &= \hat{A}_{l,a, \omega} \hat{\rho} \hat{A}_{k, a, \omega}^{\dagger} \hat{H}_S
    + \omega \hat{A}_{l,a, \omega} \hat{\rho} \hat{A}_{k, a, \omega}^{\dagger} - \omega \hat{A}_{l,a, \omega} \hat{\rho} \hat{A}_{k, a, \omega}^{\dagger} \n
    &= \hat{A}_{l,a, \omega} \hat{\rho} \hat{A}_{k, a, \omega}^{\dagger} \hat{H}_S,
\end{align}
and
\begin{align}
    \hat{H}_S \hat{A}_{k, a, \omega}^{\dagger} \hat{A}_{l,a, \omega} \hat{\rho}
    &= \hat{A}_{k, a, \omega}^{\dagger} \hat{H}_S \hat{A}_{l,a, \omega} \hat{\rho}
    + \omega \hat{A}_{k, a, \omega}^{\dagger} \hat{A}_{l,a, \omega} \hat{\rho} \n
    &= \hat{A}_{k, a, \omega}^{\dagger} \hat{A}_{l,a, \omega} \hat{H}_S \hat{\rho} 
    - \omega \hat{A}_{k, a, \omega}^{\dagger} \hat{A}_{l,a, \omega} \hat{\rho}
    + \hat{A}_{k, a, \omega}^{\dagger} \hat{A}_{l,a, \omega} \hat{\rho} \n
    &= \hat{A}_{k, a, \omega}^{\dagger} \hat{A}_{l,a, \omega} \hat{\rho} \hat{H}_S .
\end{align}
Thus, we have
$[ \hat{H}_S, \hat{A}_{l,a, \omega} \hat{\rho} \hat{A}_{k, a, \omega}^{\dagger} ] = 0$
and
$[ \hat{H}_S , \hat{A}_{k, a,\omega}^{\dagger} \hat{A}_{l, a, \omega} \hat{\rho} ] = 0$.
Therefore, $\left[ \hat{H}_S , \mathcal{L} [ \hat{\rho} ] \right] = 0$
when $[\hat{H}_S , \hat{\rho} ] = 0$,
which is the property in Eq.~(\ref{eq:Lindbladian_commute}).

Finally, we discuss upper bounds on a steady-state heat current.
For the GKSL master equation described by the Lindbladian,
we define the steady state $\hat{\rho}_{\text{s.s.}}$ from an initial state
$\hat{\rho}_S (0)$ in the following:
\begin{align}
    \hat{\rho}_{\text{ss}} &\ceq
    \lim_{t \to \infty}
    e^{\mathcal{L} t} [ \hat{\rho}_S (0) ]
    = \lim_{t \to \infty}
    \left[ \mathcal{I} + t \mathcal{L} + \frac{t^2}{2!} \mathcal{L}^2 + \cdots \right] [ \hat{\rho}_S (0) ],
\end{align}
where $\mathcal{I}$ denotes the identity operator onto the Hilbert space of the density operator.
By using the property of the Lindbladian in Eq.~(\ref{eq:Lindbladian_commute}) iteratively,
we can show the following commutation relations:
\begin{align}
    [ \hat{H}_S , \hat{\rho}_S (0)] &= 0
    \ \ \ \Rightarrow \ \ \ 
    \left[ \hat{H}_S, \mathcal{L} [ \hat{\rho}_S (0) ] \right] = 0
    \ \ \ \Rightarrow \ \ \ 
    \left[ \hat{H}_S, \mathcal{L}^2 [ \hat{\rho}_S (0) ] \right] = 0
    \ \ \ \Rightarrow \ \ \  \cdots .
\end{align}
Therefore, a steady state $\hat{\rho}_{\text{ss}}$ of the GKSL master equation from an initial state $\hat{\rho}_S (0)$ that commutes with the
system Hamiltonian, also commutes with it:
\begin{align}
    [ \hat{H}_S , \hat{\rho}_S (0) ] = 0
    \ \ \ \Rightarrow \ \ \
    [ \hat{H}_S, \hat{\rho}_{\text{ss}} ] = 0.
\end{align}

Here, let us consider a situation in which the system is coupled with several heat baths, and eventually reaches a steady state $\hat{\rho}_{\text{ss}}$.
Then, a steady-state heat current $J^{(a)}_{\text{ss}}$ flowing from a specific bath labeled by $a$ can be defined as
\begin{align}
    J^{(a)}_{\text{ss}} \ceq \text{Tr} \left[ \hat{H}_S \mathcal{D}_{\text{G}}^{(a)} [ \hat{\rho}_{\text{ss}} ] \right].
\end{align}
Then, by using the theorem in Eq.~(\ref{eq:thm1_JR_JG}),
this steady-state heat current is exactly the same as the heat current
calculated by the Redfield equation:
\begin{align}
    J_{\text{ss}}^{(a)} = J_{\text{R}}^{(a)} ( \hat{\rho}_{\text{ss}} ).
\end{align}
Therefore, we can also bound the steady-state heat current for the GKSL master equation by exploiting the already derived bound for the Redfield equation.
Moreover, as explained in the next section in detail,
we can construct a multi-bath system composed of $L$-particle system
that exhibits the steady-state
heat current $| J_{\text{ss}}^{(a)} | = \Theta (L^3)$.
This means that our derived bound for the steady-state heat current
gives us the best achievable scaling of the heat current.

\subsection{Example of quantum enhanced thermodynamic devices}
\label{subsec:GKSL_appl}

\subsubsection{Quantum heat engine}
\label{subsubsec:QHE}
\begin{figure}
	\begin{center}
		\includegraphics[clip,width=8cm,bb=0 0 525 425]{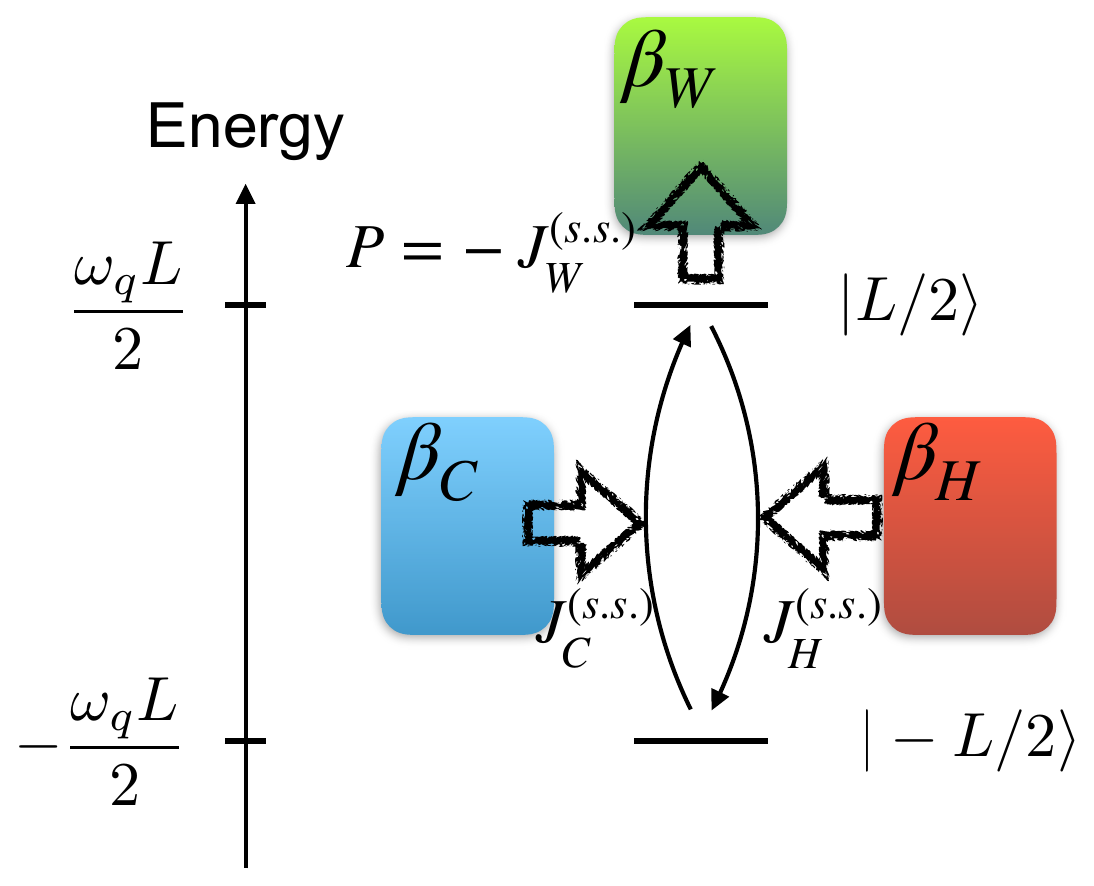}
		\caption{Schematic of quantum heat engine}
		\label{fig:QHE}
	\end{center}
\end{figure}

Utilizing the $L$-body interaction described in the main text,
we can implement a quantum heat engine that exhibits a power output proportional to $L^3$ with a fixed value of the efficiency
(See Fig.~\ref{fig:QHE} for a schematic of the heat engine).
This exceeds not only a scaling $\Theta (L)$ of the conventional heat engine with separable qubits but also
a scaling $\Theta (L^2)$ of
recently proposed heat engines based on superabsorption
with an entanglement~\cite{kamimura2022quantum}.
Moreover, the model of quantum heat engine realizes a
steady-state heat current that scales as $\Theta (L^3)$,
which allows us to saturate Bound~1 even in a steady state.

As we do in the main text, We again consider
a system Hamiltonian given by
\begin{align}
    \hat{H}_S = \omega_q \hat{J}_z.
\end{align}
We take an interaction
Hamiltonian expressed as
\begin{align}
    \hat{H}_{\text{int}} = g L \left[  \bigotimes_{i=1}^L \hat{\sigma}_x^{(i)} \right] \otimes \hat{B},
\end{align}
which satisfies the extensivity
$\| \hat{H}_{\text{int}} \| = \Theta(L)$
due to the property
$\left\| \bigotimes_{i=1}^L \hat{\sigma}_x^{(i)} \right\| = 1$.
Under a condition $\omega_q \gg g,$
we can apply a rotating-wave approximation 
and obtain a GKSL master equation.
Starting from an initial state that is diagonal in two Dicke states
$\ket{ L/2 }$ and $\ket{ -L/2 }$,
the system dynamics is
totally confined in this specific subspace,
and obey the following equation:
\begin{align}
    \frac{d}{dt}
    \begin{pmatrix}
        p_{L/2} (t) \\
        p_{-L/2} (t) 
    \end{pmatrix}
    = L^2 \sum_{a}
    \begin{pmatrix}
        - \gamma_{\downarrow}^{(a)} & \gamma_{\uparrow}^{(a)} \\
        \gamma_{\downarrow}^{(a)} & - \gamma_{\uparrow}^{(a)}
    \end{pmatrix}
    \begin{pmatrix}
        p_{L/2} (t) \\
        p_{-L/2} (t) 
    \end{pmatrix},
\end{align}
where a quantum state of the system at time $t$ is expressed as
$\hat{\rho}_S (t) = p_{L/2} (t) \ket{ L/2 } \! \bra{ L/2 } + p_{-L/2} (t) \ket{ -L/2 } \! \bra{ -L/2 }$,
and the matrix is represented by the basis
$\{ \ket{L/2} \! \bra{L/2}, \ket{- L/2} \! \bra{-L/2} \}$.
Let us consider a case where we have three baths,
respectively labeled by H (hot), C (cold) and W (work),
and let $\beta_{\text{H}}, \beta_{\text{C}}$ and $\beta_{\text{W}}$
respectively denote inverse temperatures of each bath.
Here, we consider a case in which
the three heat baths are white-noise environments,
and the inverse temperature $\beta_X$ is defined as
$\beta_X = \beta^{(0)}_X / L$
with $\beta^{(0)}_X$ being a constant about $L$
($X =$ H, C, W).
Then, the dissipation coefficients $ \gamma^{(a)}_{\downarrow} $ and $ \gamma^{(a)}_{\uparrow} $ are 
independent of $L$,
and a steady state is given as
\begin{align}
    \hat{\rho}_{\text{ss}}
    = \begin{pmatrix}
        p^{ \text{(s.s.)} }_{L/2} \\
        p^{ \text{(s.s.)} }_{-L/2}
    \end{pmatrix}
    = \frac{1}{ \Gamma_{\downarrow} + \Gamma_{\uparrow} } 
    \begin{pmatrix}
      \Gamma_{\uparrow} \\
      \Gamma_{\downarrow}
    \end{pmatrix},
\end{align}
where we introduce a summation of the dissipation coefficients as
$\Gamma_{\uparrow (\downarrow) } = \sum_{a = \text{H,C,W} } \gamma^{(a)}_{\uparrow (\downarrow) } $.
For this steady state, we interpret a heat current $- J_{\text{W}}^{(\text{s.s.})}$
flowing from the $L$-qubit system to bath W as a power output $P$,
which reads
\begin{align}
    P = - J^{\text{(s.s.)}}_{\text{W}} = \frac{ \omega_q L^3 }{ \Gamma_{\downarrow} + \Gamma_{\uparrow} }
    \left( \gamma^{\text{(W)}}_{\downarrow} \Gamma_{\uparrow} - \gamma^{\text{(W)}}_{\uparrow} \Gamma_{\downarrow} \right).
\end{align}
This scales as $P = \Theta (L^3)$.
On the other hand, from the first law of thermodynamics in a steady state,
which is expressed as
$J^{\text{(s.s.)}}_{\text{H}} + J^{\text{(s.s.)}}_{\text{C}} + J^{\text{(s.s.)}}_{\text{W}} = 0$,
the efficiency of the heat engine
$\eta^{\text{(s.s.)}} = P/ J^{\text{(s.s.)}}_{\text{H}}$ reads
\begin{align}
    \eta^{\text{(s.s.)}}
    = 1 + \frac{ J^{\text{(s.s.)}}_{\text{C}} }{ J^{\text{(s.s.)}}_{\text{H}}}
    = 1 + \frac{ \gamma^{\text{C}}_{\downarrow} \Gamma_{\uparrow} - \gamma^{\text{C}}_{\uparrow} \Gamma_{\downarrow} }{ \gamma^{\text{H}}_{\downarrow} \Gamma_{\uparrow} - \gamma^{\text{H}}_{\uparrow} \Gamma_{\downarrow} },
\end{align}
and scales as $\eta^{\text{(s.s.)}} = \Theta (L^0)$.
From the second law of thermodynamics in a steady state
$\beta_{\text{H}} J^{\text{(s.s.)}}_{\text{H}} + \beta_{\text{C}} J^{\text{(s.s.)}}_{\text{C}} + \beta_{\text{W}} J^{\text{(s.s.)}}_{\text{W}} \le 0$,
the efficiency $\eta^{\text{(s.s.)}}$ is bounded by the Carnot efficiency as $\eta^{\text{(s.s.)}} \le \eta_{\text{Carnot}} = 1 - \beta_{\text{H}} / \beta_{\text{C}}$
under the conditions of a heat engine described as
$\beta_{\text{C}} > \beta_{\text{H}} \gg \beta_{\text{W}}$,
$J^{\text{(s.s.)}}_{\text{H}} \ge 0$,
$J^{\text{(s.s.)}}_{\text{C}} \le 0$, and
$J^{\text{(s.s.)}}_{\text{W}} \le 0$.
Therefore, this heat engine exhibits a quantum-enhanced trade-off performance
as $P/ \Delta \eta = \Theta (L^3)$.
Here, we introduce an efficiency deficit as $\Delta \eta = \eta_{\text{Carnot}} - \eta^{(\text{s.s.})} \ge 0$.

\sk{On the other hand, if the condition $\beta_X = \beta^{(0)}_X / L$ is not assumed
and the temperatures are constant about $L$,
a transition from $\ket{-L/2}$ to $\ket{+L/2}$ would be exponentially suppressed by the Boltzmann factor when we increase $L$.
This is because the energy difference between the two states is proportional to $L$.}

\subsubsection{Quantum battery}
\label{subsubsec:Qbattery}
\begin{figure}
	\begin{center}
		\includegraphics[clip,width=8cm,bb=0 0 550 375]{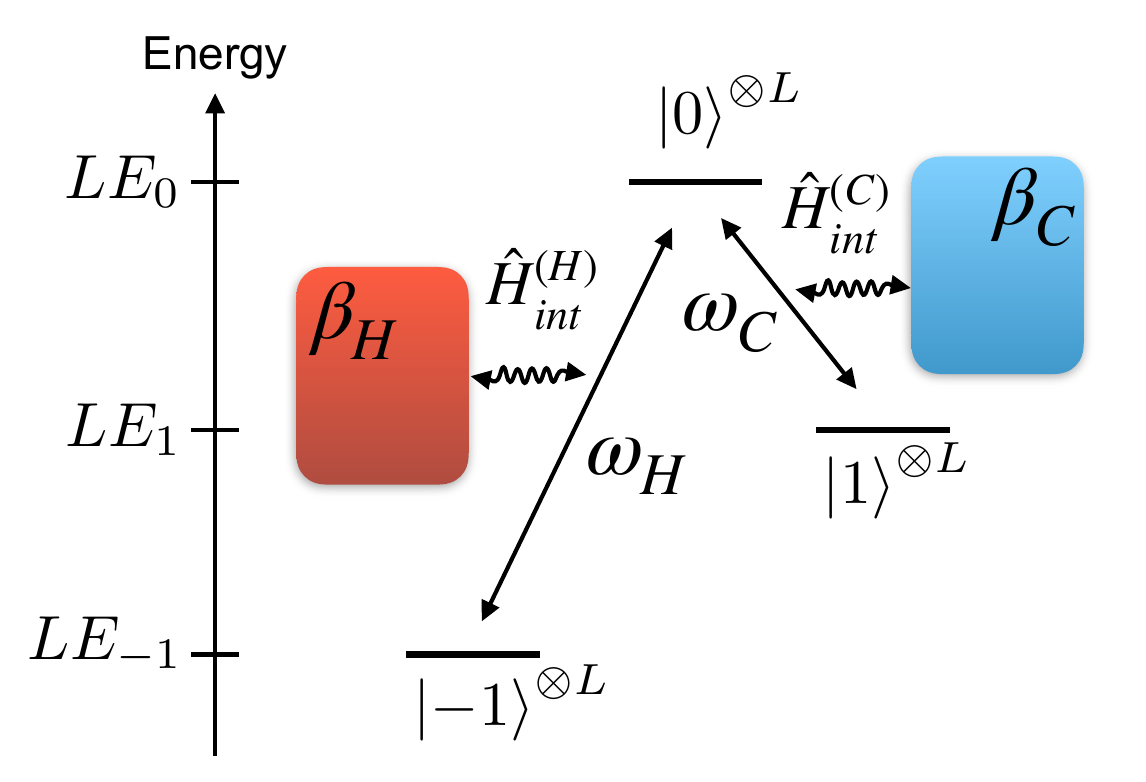}
		\caption{Schematic of quantum battery}
		\label{fig:QB}
	\end{center}
\end{figure}

We present an example of a quantum battery
composed of $L$ three-level particles
coupled with two heat baths,
labeled by H (hot) and C (cold), respectively.
(See Fig.~\ref{fig:QB} for a schematic of the quantum battery).
We define a system Hamiltonian as
\begin{align}
    \hat{H}_S = \sum_{i=1}^L \hat{h}_s^{(i)},
\end{align}
where a Hamiltonian for the $i$-th particle $\hat{h}_s^{(i)}$ reads
\begin{align}
    \hat{h}_s^{(i)} = E_1 \ket{1}_{i} \! \bra{1} + E_0 \ket{0}_{i} \! \bra{0} + E_{-1} \ket{-1}_{i} \! \bra{-1}.
\end{align}
$E_{j}$ denotes an energy of the corresponding energy eigenstate $\ket{j}_i$ of the $i$-th particle $(j = -1, 0, 1)$.
For the two heat baths H and C,
we define interaction Hamiltonians $\hat{H}^{(\text{H})}_{\text{int}}$ and $\hat{H}^{(\text{C})}_{\text{int}}$
with the $L$-particle system
as follows:
\begin{align}
    \hat{H}_{\text{int}}^{\text{(H)}} &=
    g L \left[ \hat{\sigma}_{-x}^{(1)} \otimes \hat{\sigma}_{-x}^{(2)} \otimes \cdots \otimes \hat{\sigma}_{-x}^{(L)} \right] \otimes \hat{B}^{\text{(H)}}, \\
    \hat{H}_{\text{int}}^{\text{(C)}} &=
    g L \left[ \hat{\sigma}_{+x}^{(1)} \otimes \hat{\sigma}_{+x}^{(2)} \otimes \cdots \otimes \hat{\sigma}_{+x}^{(L)} \right] \otimes \hat{B}^{\text{(C)}},
\end{align}
where $g$ is a coupling strength,
$\hat{\sigma}_{- x}^{(i)}$ and $\hat{\sigma}_{+ x}^{(i)}$ are defined as
$\hat{\sigma}_{\pm x}^{(i)} = \ket{0}_{i} \! \bra{\pm 1} + \ket{\pm 1}_{i} \! \bra{0}$,
and $\hat{B}^{(X)}$ is a bath operator for bath $X$ ($X =$ H,C).
Under these Hamiltonians, after applying the Born, Markov and rotating-wave approximations,
we obtain a GKSL master equation for the $L$-particle system.
In particular, within a subspace of quantum state
that is diagonal with respect to three energy eigenstates
$\{ \ket{1}^{\otimes L} $, $\ket{0}^{\otimes L} $, $\ket{-1}^{\otimes L} \}$,
the system dynamics are totally confined in this specific subspace,
and obey the following GKSL master equation for a quantum state
$\hat{\rho}_S (t) = \sum_{j=1,0,-1} p^{(L)}_{j} (t) \ket{j}^{\otimes L} \! \bra{j}$:
\begin{align}
    \frac{d}{dt} 
    \begin{pmatrix}
        p_{1}^{(L)} (t) \\
        p_{0}^{(L)} (t) \\
        p_{-1}^{(L)} (t) 
    \end{pmatrix}
    = L^2 \left( \bm{M}^{(\text{H})} + \bm{M}^{(\text{C})} \right)
     \begin{pmatrix}
        p_{1}^{(L)} (t) \\
        p_{0}^{(L)} (t) \\
        p_{-1}^{(L)} (t) 
    \end{pmatrix},
\end{align}
where the matrix
$\bm{M}^{(\text{H})}$ and $\bm{M}^{(\text{C})}$
are
responsible for dissipation dynamics induced by bath H and bath C, respectively.
Here,
as we do in the previous example of quantum heat engine,
we assume that the two heat baths are white-noise environments,
and we define the inverse temperature as
$\beta_X = \beta_X^{(0)} / L$,
where $\beta_X^{(0)}$ is a constant about $L$
($X = $ H, C).
Then, these two matrices
$\bm{M}^{(\text{H})}$ and $\bm{M}^{(\text{C})}$
are independent of $L$,
and this immediately means that the dynamics are $L^2$ times faster compared with those for a single particle case.
$\bm{M}^{(\text{H})}$ and $\bm{M}^{(\text{C})}$ explicitly read 
\begin{align}
    \bm{M}^{(\text{H})} &= 
    \begin{pmatrix}
        0 & 0 & 0 \\
        0 & - \gamma_{\downarrow}^{\text{(H)}} & \gamma_{\uparrow}^{\text{(H)}} \\
        0 & \gamma_{\downarrow}^{\text{(H)}} & -\gamma_{\uparrow}^{\text{(H)}}
    \end{pmatrix}, \\
     \bm{M}^{(\text{C})} &= 
    \begin{pmatrix}
        - \gamma_{\uparrow}^{\text{(C)}} & \gamma_{\downarrow}^{\text{(C)}} & 0 \\
        \gamma_{\uparrow}^{\text{(C)}} & - \gamma_{\downarrow}^{\text{(C)}} & 0 \\
         0 & 0 & 0 
    \end{pmatrix}.
\end{align}
Reflecting the definition of the interaction Hamiltonian,
bath H (C) induces a transition between the two eigenstate $\ket{0}^{\otimes L}$ and $\ket{-1}^{\otimes L}$ ($\ket{1}^{\otimes L}$),
and the energy difference of this transition is given as $\omega_{\text{H}} = L (E_0 - E_{-1})$ ($\omega_{\text{C}} = L (E_0 - E_{1}) $).
Under a condition that $E_{-1} < E_{1} < E_0$,
we have a detailed-balance condition $\gamma_{\downarrow}^{(X)} / \gamma_{\uparrow}^{(X)} = e^{ \beta_X \omega_X}$ ($X = $ H, C),
and a steady-state is given as
\begin{align}
   \hat{\rho}_{\text{ss}} = 
    \frac{1}{ 1 + e^{ \beta_{\text{C}} \omega_{\text{C}} } + e^{ \beta_{\text{H}} \omega_{\text{H}} } }
    \begin{pmatrix}
        e^{ \beta_{\text{C}} \omega_{\text{C}} } \\
        1 \\
        e^{ \beta_{\text{H}} \omega_{\text{H}} }
    \end{pmatrix}.
\end{align}
Under an additional condition
$\beta_{\text{H}} \omega_{\text{H}} < \beta_{\text{C}} \omega_{\text{C}}$,
we obtain a population inversion between the eigenstates $\ket{1}^{\otimes L}$ and $\ket{-1}^{\otimes L}$.
Then, an ergotropy $\mathcal{E}_L^{\text{(s.s.)}}$ for the steady-state,
which is the maximally exractable energy from the system,
is evaluated as follows~\cite{lenard1978thermodynamical}:
\begin{align}
    \mathcal{E}_L^{\text{(s.s.)}} = \text{Tr} [ \hat{H}_S \hat{\rho}_S^{\text{(s.s.)}} ] - \min_{ \hat{U} \text{: unitary} } \text{Tr} [ \hat{H}_S \hat{U} \hat{\rho}_S^{\text{(s.s.)}} \hat{U}^{\dagger} ]
    = \frac{ L (e^{ \beta_{\text{C}} \omega_{\text{C}} } - e^{ \beta_{\text{H}} \omega_{\text{H}} } ) }{ 1 + e^{ \beta_{\text{C}} \omega_{\text{C}} } + e^{ \beta_{\text{H}} \omega_{\text{H}} } } (E_1 - E_{-1} ) > 0,
\end{align}
where ``min" is taken for an arbitrary unitary operation $\hat{U}$ acting on the system.
Consequently, for both of a collective charging process
and a parallel one
(where $L$ particles are charged independently),
the same amount of ergotropy $\mathcal{E}_L^{\text{(s.s.)}} = L \mathcal{E}_{L=1}^{\text{(s.s.)}}$
is stored in a steady-state,
while a charging speed for the former is $L^2$ times faster than that for the latter
when both of the systems start from the same initial state $\ket{-1}^{\otimes L}$.
This directly means that a {\it quantum advantage} $\Gamma_{\text{adv}}$,
which quantifies a ratio of charging powers for the collective scheme and parallel one,
scales as $\Gamma_{\text{adv}} = \Theta (L^2)$.
That exhibits a {\it superextensive} scaling of the advantage,
which is prohibited for an isolated quantum battery~\cite{gyhm2022quantum}.

\section{Experimentally feasible system for demonstration of Bound~1}
\label{sec:experiment}
In this section, we explain how to realize our proposal by using superconducting qubits.
Let us consider a three-qubit system where we use one of them as an ancillary qubit to induce a decay process.
It is known that we can realize a three-body interaction by using superconducting qubits~\cite{pedersen2019native},
and we adopt such a system.
The Hamiltonian is as follows:
\begin{align}
    \hat{H} (t)
    &= \hat{H}_{\text{3-qubit}} + \hat{H}_{\text{drv}} (t) \\
    &= \frac{\Omega_q}{2} \left( \hat{\sigma}_z^{(1)} + \hat{\sigma}_z^{(2)} \right) + \frac{\omega_a}{2} \hat{\sigma}_z^{(a)}
    - J_z^{(1,2)} \hat{\sigma}_z^{(1)} \otimes \hat{\sigma}_z^{(2)}
    - J_z^{(a)} \left( \hat{\sigma}_z^{(1)} + \hat{\sigma}_z^{(2)} \right) \otimes \hat{\sigma}_z^{(a)} \n
    & \ \ \ \ \ + J_z \hat{\sigma}_z^{(1)} \otimes \hat{\sigma}_z^{(2)} \otimes \frac{ 1 - \hat{\sigma}_z^{(a)} }{2}
    + J_x \hat{\sigma}_x^{(1)} \otimes \hat{\sigma}_x^{(2)} \otimes \frac{ 1 - \hat{\sigma}_z^{(a)} }{2} \n
    & \ \ \ \ \  + \lambda \hat{\sigma}_x^{(a)} \cos (\omega_a t).
\end{align}
where $\hat{H}_{\text{drv}} (t) = \lambda \hat{\sigma}_x^{(a)} \cos (\omega_a t)$
is a driving term on the ancillary qubit.
$\Omega_q$ denotes the energy of the first and second qubits,
 $\omega_a$ is the energy of the ancillary qubit,
all $J$'s are the coupling strengths, and 
$\lambda$ describes the amplitude of the driving on the ancillary qubit.
We assume that these parameters are positive.
Moving into a rotating frame with a unitary operator
$\hat{U}_1 (t) = e^{ \frac{ i \omega_a t}{2} \hat{\sigma}_z^{(a)} }$,
we 
obtain the Hamiltonian $\hat{H}_1 (t)$ in this frame as follows:
\begin{align}
    \hat{H}_1 (t)
    &= \hat{U}_1 (t) \hat{H} (t) \hat{U}_1^{\dagger} (t)
    + i \frac{ \pd \hat{U}_1 (t) }{ \pd t } \hat{U}_1^{\dagger} (t) \n
    &= \hat{H}_{\text{3-qubit}} + \lambda \cos (\omega_a t) e^{ \frac{ i \omega_a t}{2} \hat{\sigma}_z^{(a)} } \hat{\sigma}_x^{(a)} e^{ \frac{ - i \omega_a t}{2} \hat{\sigma}_z^{(a)} }
    - \frac{\omega_a}{2} \hat{\sigma}_z^{(a)} \n
    &= \hat{H}_{\text{3-qubit}} - \frac{\omega_a}{2} \hat{\sigma}_z^{(a)}
    + \lambda \frac{ e^{ i \omega_a t} + e^{-i \omega_a t} }{2}
    \left[ \frac{ e^{ i \omega_a t} + e^{-i \omega_a t} }{2} \hat{\sigma}_x^{(a)} + i \frac{ e^{ i \omega_a t} - e^{-i \omega_a t} }{2} \hat{\sigma}_y^{(a)} \right] \n
    &= \frac{\Omega_q}{2} \left( \hat{\sigma}_z^{(1)} + \hat{\sigma}_z^{(2)} \right)
    - J_z^{(1,2)} \hat{\sigma}_z^{(1)} \otimes \hat{\sigma}_z^{(2)}
    - J_z^{(a)} \left( \hat{\sigma}_z^{(1)} + \hat{\sigma}_z^{(2)} \right) \otimes \hat{\sigma}_z^{(a)} \n
    & \ \ \ \ \  + J_z \hat{\sigma}_z^{(1)} \otimes \hat{\sigma}_z^{(2)} \otimes \frac{ 1 - \hat{\sigma}_z^{(a)} }{2}
    + J_x \hat{\sigma}_x^{(1)} \otimes \hat{\sigma}_x^{(2)} \otimes \frac{ 1 - \hat{\sigma}_z^{(a)} }{2} \n
    & \ \ \ \ \ + \frac{\lambda}{2} \left[ \hat{\sigma}_+^{(a)} + \hat{\sigma}_-^{(a)} \right]
    + \frac{\lambda}{2} \left[ e^{ +2 i \omega_a t} \hat{\sigma}_+^{(a)} + e^{ - 2 i \omega_a t} \hat{\sigma}_-^{(a)} \right].
\end{align}
Next, we change the
notation of the Pauli matrices of the ancillary qubit as follows:
\begin{align}
    \hat{\sigma}_x^{(a)} &\mapsto \hat{\sigma}_z^{(a)}, &
    \hat{\sigma}_y^{(a)} &\mapsto \hat{\sigma}_y^{(a)}, &
    \hat{\sigma}_z^{(a)} &\mapsto -\hat{\sigma}_x^{(a)}. &
\end{align}
Then, the lowering and raising operators of the ancillary qubit are respectively changed as
\begin{align}
    \hat{\sigma}_{\pm}^{(a)} = \frac{1}{2} \left( \hat{\sigma}_x^{(a)} \pm i \hat{\sigma}_y^{(a)} \right)
    \mapsto \frac{1}{2} \left( \hat{\sigma}_z^{(a)} \pm i \hat{\sigma}_y^{(a)} \right),
\end{align}
and the Hamiltonian is changed as
\begin{align}
    \hat{H}_1 (t) \mapsto \hat{H}_2 (t)
    &= \frac{\Omega_q}{2} \left( \hat{\sigma}_z^{(1)} + \hat{\sigma}_z^{(2)} \right)
    + \frac{\lambda}{2} \hat{\sigma}_z^{(a)} 
    + \left[ \frac{1}{2} J_z - J_z^{(1,2)} \right]  \hat{\sigma}_z^{(1)} \otimes \hat{\sigma}_z^{(2)}
    + \frac{1}{2} J_x \hat{\sigma}_x^{(1)} \otimes \hat{\sigma}_x^{(2)} \n
    & \ \ \ \ \ + J_z^{(a)} \left( \hat{\sigma}_z^{(1)} + \hat{\sigma}_z^{(2)} \right) \otimes \hat{\sigma}_x^{(a)}
    +\frac{1}{2} J_z \hat{\sigma}_z^{(1)} \otimes \hat{\sigma}_z^{(2)} \otimes \hat{\sigma}_x^{(a)}
    +\frac{1}{2} J_x \hat{\sigma}_x^{(1)} \otimes \hat{\sigma}_x^{(2)} \otimes \hat{\sigma}_x^{(a)} \n
    & \ \ \ \ \ +\frac{\lambda}{4} \left[ e^{ +2 i \omega_a t} \left( \hat{\sigma}_z^{(a)} + i \hat{\sigma}_y^{(a)} \right)
    + e^{ -2 i \omega_a t} \left( \hat{\sigma}_z^{(a)} - i \hat{\sigma}_y^{(a)} \right) \right].
\end{align}
Let us define $\chi \ceq J_z - 2 J_z^{(1,2)}$ as an
Ising-type coupling strength between the first and second qubit.
The Hamiltonian can be rewritten as
\begin{align}
    \hat{H}_2 (t) 
    &= \frac{\Omega_q}{2} \left( \hat{\sigma}_z^{(1)} + \hat{\sigma}_z^{(2)} \right)
    + \frac{\lambda}{2} \hat{\sigma}_z^{(a)} 
    + \frac{\chi}{2} \hat{\sigma}_z^{(1)} \otimes \hat{\sigma}_z^{(2)}
    + \frac{1}{2} J_x \hat{\sigma}_x^{(1)} \otimes \hat{\sigma}_x^{(2)} \n
    & \ \ \ \ \ + J_z^{(a)} \left( \hat{\sigma}_z^{(1)} + \hat{\sigma}_z^{(2)} \right) \otimes \hat{\sigma}_x^{(a)}
    +\frac{1}{2} J_z \hat{\sigma}_z^{(1)} \otimes \hat{\sigma}_z^{(2)} \otimes \hat{\sigma}_x^{(a)}
    +\frac{1}{2} J_x \hat{\sigma}_x^{(1)} \otimes \hat{\sigma}_x^{(2)} \otimes \hat{\sigma}_x^{(a)} \n
    & \ \ \ \ \ +\frac{\lambda}{4} \left[ \left( e^{ +2 i \omega_a t} + e^{ -2 i \omega_a t} \right) \hat{\sigma}_z^{(a)} + 
    \left( e^{ +2 i \omega_a t} - e^{ -2 i \omega_a t} \right) 
    \left( \hat{\sigma}_+^{(a)} - \hat{\sigma}_-^{(a)} \right) \right],
\end{align}
where we use $ \hat{\sigma}_y^{(a)} = -i \left( \hat{\sigma}^{(a)}_+ - \hat{\sigma}^{(a)}_- \right)$.
Here, we move into another rotating frame according to the unitary operator $\hat{U}_3 (t)$
that is defined as
\begin{align}
    \hat{U}_3 (t) \ceq
    e^{+ \frac{i \lambda t}{2} \hat{\sigma}_z^{(a)} }
    e^{+ \frac{i \Omega_q t}{2} \hat{\sigma}_z^{(2)} }
    e^{+ \frac{i \Omega_q t}{2} \hat{\sigma}_z^{(1)} }.
\end{align}
Then, the Hamiltonian is changed as
\begin{align}
    \hat{H}_2 (t) \mapsto 
    \hat{H}_3 (t)
    &\ceq \hat{U}_3 (t) \hat{H}_2 (t) \hat{U}_3^{\dagger} (t)
    + i \frac{ \pd \hat{U}_3 (t) }{ \pd t } \hat{U}_3^{\dagger} (t) \n
    &= \frac{\chi}{2} \hat{\sigma}_z^{(1)} \otimes \hat{\sigma}_z^{(2)}
    + \frac{1}{2} J_x \left[ e^{i \Omega_q t} \hat{\sigma}_+^{(1)} +  e^{ -i \Omega_q t} \hat{\sigma}_-^{(1)} \right] \otimes
    \left[ e^{i \Omega_q t} \hat{\sigma}_+^{(2)} +  e^{ -i \Omega_q t} \hat{\sigma}_-^{(2)} \right] \n
    & \ \ \ \ \ + J_z^{(a)} \left( \hat{\sigma}_z^{(1)} + \hat{\sigma}_z^{(2)} \right) \otimes
     \left[ e^{i \lambda t} \hat{\sigma}_+^{(a)} + e^{- i \lambda t} \hat{\sigma}_-^{(a)} \right]
    +\frac{1}{2} J_z \hat{\sigma}_z^{(1)} \otimes \hat{\sigma}_z^{(2)} \otimes
    \left[ e^{i \lambda t} \hat{\sigma}_+^{(a)} + e^{- i \lambda t} \hat{\sigma}_-^{(a)} \right] \n
    & \ \ \ \ \ +\frac{1}{2} J_x
    \left[ e^{i \Omega_q t} \hat{\sigma}_+^{(1)} +  e^{ -i \Omega_q t} \hat{\sigma}_-^{(1)} \right] \otimes
    \left[ e^{i \Omega_q t} \hat{\sigma}_+^{(2)} +  e^{ -i \Omega_q t} \hat{\sigma}_-^{(2)} \right] \otimes
    \left[ e^{i \lambda t} \hat{\sigma}_+^{(a)} +  e^{ -i \lambda t} \hat{\sigma}_-^{(a)} \right] \n
    & \ \ \ \ \ +\frac{\lambda}{4} \left[ \left( e^{ +2 i \omega_a t} + e^{ -2 i \omega_a t} \right) \hat{\sigma}_z^{(a)} + 
    \left( e^{ +2 i \omega_a t} - e^{ -2 i \omega_a t} \right) 
    \left( e^{i \lambda t} \hat{\sigma}_+^{(a)} - e^{- i \lambda t} \hat{\sigma}_-^{(a)} \right) \right].
\end{align}
Here, we apply a rotating-wave approximation
to the Hamiltonian $\hat{H}_3 (t)$.
Namely, we ignore the fast-oscillating terms having non-zero
frequencies under the following conditions:
\begin{align}
    J_x, J_z^{(a)}, J_z \ll \Omega_q, &&
    \lambda = 2 \Omega_q, &&
    \lambda \ll \omega_a.
\end{align}
Then, we have the following Hamiltonian $\hat{H}_{\text{RWA},3}$ 
after the rotating-wave approximation:
\begin{align}
    \hat{H}_3 (t)
    \simeq \hat{H}_{\text{RWA},3}
    &\ceq \frac{\chi}{2} \hat{\sigma}_z^{(1)} \otimes \hat{\sigma}_z^{(2)}
    + \frac{1}{2} J_x \left( \hat{\sigma}_+^{(1)} \otimes \hat{\sigma}_-^{(2)} + \text{h.c.} \right)
    + \frac{1}{2} J_x \left( \hat{\sigma}_+^{(1)} \otimes \hat{\sigma}_+^{(2)} \otimes \hat{\sigma}_-^{(a)}
    + \text{h.c.} \right).
\end{align}
Then, going
to the
Schr\"odinger picture,
we have the following total Hamiltonian:
\begin{align}
    \hat{H}_{\text{RWA}}
    &= \hat{H}_{\text{RWA,3}} + \frac{\Omega_q}{2} \left( \hat{\sigma}_z^{(1)} + \hat{\sigma}_z^{(2)} \right)
    + \frac{\lambda}{2} \hat{\sigma}_z^{(a)}
    = \hat{H}_{\text{2-qubit}} + \hat{H}_{\text{int}} + \hat{H}_{\text{ancillary}},
\end{align}
where we 
have
\begin{align}
    \hat{H}_{\text{2-qubit}} &\ceq \frac{\Omega_q}{2} \left( \hat{\sigma}_z^{(1)} + \hat{\sigma}_z^{(2)} \right)
    + \frac{\chi}{2} \hat{\sigma}_z^{(1)} \otimes \hat{\sigma}_z^{(2)}
    + \frac{1}{2} J_x \left( \hat{\sigma}_+^{(1)} \otimes \hat{\sigma}_-^{(2)} + \text{h.c.} \right), \n
    \hat{H}_{\text{int}} &\ceq \frac{1}{2} J_x \left( \hat{\sigma}_+^{(1)} \otimes \hat{\sigma}_+^{(2)} \otimes \hat{\sigma}_-^{(a)}
    + \text{h.c.} \right), \n
    \hat{H}_{\text{ancillary}} &\ceq \frac{\lambda}{2} \hat{\sigma}_z^{(a)} .
\end{align}
Let us consider a case in which the initial state of the first and second qubit is prepared in a subspace spanned by $|00\rangle $ and $|11\rangle $.
This subspace is decoupled from the other subspace spanned by $ \frac{1}{\sqrt{2}} ( \ket{01} \pm \ket{10} )$,
and thus we can ignore the second and third terms in
$\hat{H}_{\text{2-qubit}}$.
Then, we can treat the two-qubit system as an effective two-level system,
and let us define the logical qubit as follows:
\begin{align}
    \ket{\tilde{0}} &\ceq \ket{00}, &
    \ket{\tilde{1}} &\ceq \ket{11}.
\end{align}
Then, we can define an effective Hamiltonian $\hat{H}_{\text{eff}}$ as follows:
\begin{align}
    \hat{H}_{\text{eff}} \ceq \Omega_q \tilde{\sigma}_z
    + \frac{1}{2} J_x \left( \tilde{\sigma}_+ \otimes \hat{\sigma}_-^{(a)} + \tilde{\sigma}_- \otimes \hat{\sigma}_+^{(a)} \right)
    + \frac{\lambda}{2} \hat{\sigma}_z^{(a)},
\end{align}
where we introduced the effective ``Pauli" matrices acting on the two-qubit
degrees of freedom in the subspace, as
\begin{align}
    \tilde{\sigma}_z &\ceq \ket{\tilde{0}} \! \bra{\tilde{0}} - \ket{\tilde{1}} \! \bra{\tilde{1}}, &
    \tilde{\sigma}_+ &\ceq \ket{\tilde{0}} \! \bra{\tilde{1}}, &
    \tilde{\sigma}_- &\ceq \ket{\tilde{1}} \! \bra{\tilde{0}}.
\end{align}

Let us consider the case in which
the ancillary qubit is strongly coupled to a transmission line
in a low temperature compared with the ancillary-qubit frequency.
Then, the dynamics $\hat{\rho}_{\text{3-qubit}} (t) $ of the total
system
obeys
the following master equation including a dissipative part on the ancillary qubit:
\begin{align}
    \frac{d}{dt} \hat{\rho}_{\text{3-qubit}} (t)
    = -i [ \hat{H}_{\text{eff}} , \hat{\rho}_{\text{3-qubit}} (t) ] + \mathcal{D}_{\text{ancillary}} [ \hat{\rho}_{\text{3-qubit}} (t) ],
\end{align}
where the dissipative part is defined as
\begin{align}
    \mathcal{D}_{\text{ancillary}} [ \hat{\rho} ] \ceq \kappa
    \left( \hat{\sigma}_-^{(a)} \hat{\rho} \hat{\sigma}_+^{(a)} 
    - \frac{1}{2} \{ \hat{\sigma}_+^{(a)} \hat{\sigma}_-^{(a)} , \hat{\rho}  \} \right).
\end{align}
If the ancillary qubit decays into its ground state
much faster than the typical time scale of dynamics induced by the interaction
between the two-qubit system and the ancillary qubit,
i.e. $ \kappa \gg J_x$, then we can adiabatically eliminate the ancillary-qubit,
and obtain the following GKSL master equation only for
the
density operator
$\hat{\rho} (t)$
for the two-qubit system:
\begin{align}
    \frac{d}{dt}\hat{\rho} (t)
    = -i [ \hat{H}_{\text{eff}} , \hat{\rho} (t)] + \frac{J_x^2}{\kappa}
    \left( \hat{L} \hat{\rho}(t) \hat{L}^{\dagger}
    - \frac{1}{2} \{ \hat{L}^{\dagger} \hat{L}, \hat{\rho}(t) \} \right),
\end{align}
where we define the Lindblad operator as $\hat{L} \ceq \hat{\sigma}_-^{(1)} \otimes \hat{\sigma}_-^{(2)}$,
and the modified decay rate $ J_x^2 / \kappa$ describes the
effective decay rate~\cite{butler2011polarization,purcell1946resonance,wood2014cavity,bienfait2016reaching,agarwal1971brownian}.
This is the realization of our proposal explained in Sec.~III~B for $L=2$.
For this system, from the transmission line,
we should be able to measure microwave photons~\cite{houck2007generating} that correspond to the heat current.
Thus, our proposal is feasible by using superconducting qubits.
Moreover, if the ancillary qubit is coupled with three transmission lines, we can realize a heat engine or a refrigerator.
For the case of the refrigerator, we can use this system to decrease the temperature of the transmission line as we explained in Sec.~IV~D, which is practically useful for quantum information processing.
\bibliographystyle{apsrev4-1}
\bibliography{QHEnogo_suppl}